%% file: Main.tex
\documentclass[acmsmall]{acmart}

\usepackage[T1]{fontenc} 
\usepackage[utf8]{inputenc}
\usepackage{amsmath}
\usepackage{bm} 
\usepackage{graphicx}
\usepackage{tikz}
\usepackage{tcolorbox}
\usepackage{pdflscape}
\usepackage{rotating}
\usepackage{fancyhdr} 
\usepackage{color, colortbl}
\usepackage{longtable}
\usepackage{multirow}
\usepackage{graphicx}
\usepackage{enumitem}
\usepackage{fontawesome}
\usepackage{multirow}
\usepackage{balance}
\usepackage{tablefootnote}
\usepackage{colortbl}
\tcbuselibrary{skins}
\usepackage{hyperref}
\usepackage[export]{adjustbox}
\usepackage{framed}
\usepackage{arydshln}
\newcolumntype{M}[1]{>{\centering\arraybackslash}m{#1}}
\newcolumntype{N}{@{}m{0pt}@{}}

\hypersetup
{
 colorlinks = true, 
 urlcolor = blue, 
 linkcolor = blue, 
 citecolor = blue 
}

\usepackage{graphicx}
\usepackage{svg}
\usepackage{subfig}
\usepackage{array}
\definecolor{mybeige}{HTML}{FFF7F3}
\definecolor{myoffwhite}{HTML}{F1F1F1}
\definecolor{mydarkpurple}{HTML}{49006A}
\definecolor{mypurple}{HTML}{99017B}
\definecolor{mydarkpink}{HTML}{E23E99}
\definecolor{mypink}{HTML}{F767A1}
\definecolor{mypink2}{HTML}{F769A1}
\definecolor{mylightpink}{HTML}{F994B1}
\definecolor{mysalmon}{HTML}{FCC8C3}
\definecolor{mylightsalmon}{HTML}{FBBABD}

\AtBeginDocument{%
  }

\setcopyright{acmlicensed}
\copyrightyear{2024}
\acmYear{2024}
\acmDOI{XXXXXXX.XXXXXXX}

\acmJournal{JACM}
\acmVolume{0}
\acmNumber{0}
\acmArticle{0}
\acmMonth{0}

\begin{document}

\title{Fairness Concerns in App Reviews: A Study on AI-based Mobile Apps}

\author{Ali Rezaei Nasab}
\email{rezaei.ali.nasab@gmail.com}
\author{Maedeh Dashti}
\email{maedeh.dashti.research@gmail.com}
\affiliation{%
  \institution{School of Computer Science, Wuhan University}
  \city{Wuhan}
  \country{China}
}



\author{Mojtaba Shahin}
\affiliation{%
  \institution{RMIT University}
  \city{Melbourne}
  \country{Australia}}
\email{mojtaba.shahin@rmit.edu.au}

\author{Mansooreh Zahedi}
\affiliation{%
  \institution{University of Melbourne}
  \city{Melbourne}
  \country{Australia}}
\email{mansooreh.zahedi@unimelb.edu.au}

\author{Hourieh Khalajzadeh}
\affiliation{%
 \institution{Deakin University}
 \city{Melbourne}
 \country{Australia}}
\email{hkhalajzadeh@deakin.edu.au}

\author{Chetan Arora}
\affiliation{%
  \institution{Monash University}
  \city{Melbourne}
  \country{Australia}}
\email{chetan.arora@monash.edu}

\author{Peng Liang}
\authornote{Corresponding author}
\affiliation{%
  \institution{School of Computer Science, Wuhan University}
  \city{Wuhan}
  \country{China}
  }
\email{liangp@whu.edu.cn}

\renewcommand{\shortauthors}{Rezaei Nasab et al.}

\begin{abstract}
Fairness is one of the socio-technical concerns that must be addressed in software systems. Considering the popularity of mobile software applications (apps) among a wide range of individuals worldwide, mobile apps with unfair behaviors and outcomes can affect a significant proportion of the global population, potentially more than any other type of software system. Users express a wide range of socio-technical concerns in mobile app reviews. This research aims to investigate fairness concerns raised in mobile app reviews. Our research focuses on AI-based mobile app reviews as the chance of unfair behaviors and outcomes in AI-based mobile apps may be higher than in non-AI-based apps. To this end, we first manually constructed a ground-truth dataset, including 1,132 fairness and 1,473 non-fairness reviews. Leveraging the ground-truth dataset, we developed and evaluated a set of machine learning and deep learning models that distinguish fairness reviews from non-fairness reviews. Our experiments show that our best-performing model can detect fairness reviews with a precision of 94\%. We then applied the best-performing model on approximately 9.5M reviews collected from 108 AI-based apps and identified around 92K fairness reviews. Next, applying the K-means clustering technique to the 92K fairness reviews, followed by manual analysis, led to the identification of six distinct types of fairness concerns (e.g., `\textit{receiving different quality of features and services in different platforms and devices}' and `\textit{lack of transparency and fairness in dealing with user-generated content}'). Finally, the manual analysis of 2,248 app owners' responses to the fairness reviews identified six root causes (e.g., ‘copyright issues’) that app owners report to justify fairness concerns.
\end{abstract}





\maketitle

\section{Introduction}
The global AI market size, particularly in the form of Machine Learning (ML) and Deep Learning (DL), is predicted to exceed USD~1.6 trillion by 2030~\cite{marketsizeAI}. The applications of AI technologies in our modern life and society vary, ranging from detecting diseases to predicting traffic jams, from enabling personalized shopping to identifying important targets in combat environments~\cite{mehrabi2021survey}. Despite the wide adoption of AI solutions, a growing concern exists about ensuring the responsible behavior of AI solutions~\cite{mikalef2022thinking}. The increasing need for responsible AI systems arises from the possibility that they could produce biased results, have errors, or lack sufficient transparency~\cite{trocin2021responsible}. Hence, organizations and governments worldwide have started developing standards, frameworks, and guidelines on the responsible use and adoption of AI \cite{dignum2022responsible,resaip,whatisai,IEEE_Ethics_Standard}. Fairness is one of the important principles in the existing ethical and responsible AI frameworks and standards. For example, Microsoft's Responsible AI framework includes fairness as one of its six core principles, along with reliability and safety, privacy and security, inclusiveness, transparency, and accountability \cite{whatisai}. Similarly, Google emphasizes fairness in its responsible AI practices to ensure that AI systems are developed to enhance people's lives in a fair manner \cite{resaip}. There is no single definition of fairness in AI systems, and varies in different domains~\cite{ekstrand2022fairness}. In AI-based algorithmic decision-making, fairness is defined as ``\textit{the absence of any prejudice or favoritism towards an individual or group based on their inherent or acquired characteristics}''~\cite{mehrabi2021survey}. Others have a broader view and see achieving fairness in AI systems as a socio-technical challenge~\cite{ekstrand2022fairness,selbst2019fairness,bird2020fairlearn,lu2022towards}. This means that fairness concerns in AI systems cannot only stem from technical components and data but can also be attributed to the process of building AI systems, humans, and governance (e.g., decisions made by developers or policies adopted by providers). However, a growing number of software systems leveraging AI solutions (AI-based software systems) have been accused of being unfair and producing biased and discriminatory outcomes, causing some (major) difficulties for individuals, society, and businesses \cite{mehrabi2021survey,galhotra2017fairness}.

The growth and popularity of AI have also triggered a wave of mobile software applications utilizing AI technologies (e.g., ML/DL technologies). Such a wave has been further energized by the emergence of mobile AI frameworks and libraries such as TensorFlow Lite that enable executing AI tasks solely on smartphones~\cite{xu2019first}. \textcolor{black}{Inspired by \cite{li2022ai, xu2019first}, we define AI-based mobile applications (apps) as mobile apps that utilize AI frameworks, including ML-based, DL-based, and AI service-based frameworks (e.g., TensorFlow Lite \cite{TensorFlowlite}, Google AI \cite{GoogleAI}).} With 6.64 billion smartphone users worldwide (83.32\% of the current world population)~\cite{smartphonenumber}, mobile software apps are the most prevailing type of software system used by a wide range of individuals and groups with different characteristics (e.g., age, education level, cultural background, race, and gender)~\cite{sarker2019context,fazzini2022characterizing,khalajzadeh2022supporting}. \textcolor{black}{While unfair behaviours and outcomes can manifest in any mobile app, including those not powered by AI, we posit that the chance of unfair behaviours and outcomes in AI-based mobile apps may be higher than in non-AI-based apps~\cite{mehrabi2021survey,mikalef2022thinking,lewicki2023out}.} \textcolor{black}{This increased risk is attributed to the complex nature of AI technologies, which can inadvertently encode and perpetuate biases present in their training data or opaque decision-making algorithms~\cite{mehrabi2021survey,belenguer2022ai, lepri2018fair}}.
Moreover, considering the pervasiveness of mobile apps, unfair behaviours and outcomes in (AI-based) mobile apps have far-reaching implications, affecting a significant proportion of the global population. Lastly, to the best of our knowledge, no work has focused on providing an accurate and fine-grained view of different types of fairness concerns and their possible root causes discussed in AI-based app reviews. Having said that, our work in this paper sets the stage for future research, advocating for a broader application of our dataset curation approach and classifier (discussed later in the paper) to explore and mitigate unfair behaviors in the wider landscape of mobile apps, AI-based or otherwise.

Users share a variety of technical and ethical concerns, ranging from bugs to privacy issues in mobile app reviews that reflect their experiences and opinions about the app. This research aims to investigate fairness concerns raised in mobile app reviews. We selected AI-based mobile apps as the subjects of our study for the reasons outlined earlier. To this end, we first collected $\approx$9.5M app reviews from 108 Android AI-based apps from the Google Play Store. \textcolor{black}{We then manually constructed a dataset (ground truth) composed of 1,132 fairness and 1,473 non-fairness reviews. We define a \textit{fairness review} as a review in which users express that they are treated differently, whether intentionally or unintentionally, due to their inherent, acquired, or context-specific attributes (e.g., age, language, location, gender) by AI-based apps or the decisions and policies of the AI-based app owners. Additionally, a fairness review can reflect users’ perceptions that the outcomes of AI-based apps, or the decisions and policies of the app owners, are inconsistent, non-transparent, and unreasonable.} Fig.~\ref{fig:examplereview} shows an example of a fairness review and the response made by the app owner to justify the raised fairness concern. Next, we evaluated a set of ML and DL models against the ground-truth dataset, which aims to distinguish fairness reviews from non-fairness reviews. Our experiments show that our best-performing model can detect fairness reviews with a precision of 94\%. Although the models are promising in identifying fairness reviews, they do not provide a fine-grained view of different fairness concerns. Hence, we developed and experimented with the K-means clustering technique, followed by a manual analysis to cluster and summarize fairness reviews. Our study shows that users discuss the top six fairness concerns in AI-based app reviews: \textit{`receiving different quality of features and services in different platforms and devices'}, \textit{`feeling linguistic discrimination}, \textit{`lack of transparency and fairness in dealing
with user-generated content'}, \textit{`feeling gender and racial discrimination'}, \textit{`feeling biased censorship and promotion'}, and \textit{`unfair and non-transparent advertisement and subscription policies'}. The app owners can use our approach in practice to automatically delineate key fairness-related concerns (and any underlying reasons) in their apps, combining our model and the clustering approach. To understand the underlying reasons for fairness concerns, we manually analyzed 2,248 responses made by app owners to the fairness reviews. This led to the identification of six root causes of the fairness concerns: ‘copyright issues’, ‘development complexity’, ‘buggy code’, ‘external factors’, ‘development cost’, and ‘user usage and awareness’. The primary contributions of this paper are:
\begin{itemize}
    \item We are the first to systematically investigate fairness concerns in AI-based app reviews at scale from the users' perspective.
    \textcolor{black}{\item We manually construct a ground-truth dataset consisting of 1,132 fairness reviews and 1,473 non-fairness reviews.}
    \item We develop ML and DL models that accurately distinguish fairness from non-fairness reviews.
    \item We develop a clustering and summarization technique to discover six fine-grained fairness concerns of users.
    \item We identify six root causes of fairness concerns from the app owners' perspective.
    \item We make the source code and experimental data available in our replication package~\cite{replicationpackage}.
\end{itemize}

\textbf{Paper Organization.} \textcolor{black}{Section~\ref{sec:relatedwork} summarizes the related literature. Section~\ref{sec:rqdataset} introduces our research questions and describes our dataset curation process with fairness and non-fairness reviews. Section~\ref{sec:fairnessclassifier} elaborates on the experiments with a set of models that detect fairness reviews in app reviews. Section~\ref{sec:cluster} applies the K-means clustering technique on fairness reviews to identify fairness concerns. We report the root causes of fairness concerns in Section \ref{sec:rootcause}. Section~\ref{sec:discussion} discusses the main findings. Section~\ref{sec:validity} outlines the threats to validity, and Section~\ref{sec:conclusion} concludes this work.}
\begin{figure}
\begin{center}
\noindent\resizebox{\columnwidth}{!}{\fbox{
\parbox{\columnwidth}{%
\faThumbsODown{} \textit{\textbf{User}: You should make it so that Android users can have the same emojis as iPhone users because the Android users' emojis look like crackheads.}
\newline

\faReply{} \textit{\textbf{App Owner}: Thanks for your feedback! It is mostly because of the copyright issues, so we currently use Android version of emojis. Hope the explanation helps. We'll keep improving the product!}}
}}
  

\end{center}
\vspace*{-1em}
    \caption{An example of a fairness review and the response made by the app owner to justify the raised fairness concern}
    \label{fig:examplereview}
    \vspace*{-1.5em}
\end{figure}

\section{Related Work}\label{sec:relatedwork}
In this section, we position our work in the existing literature on fairness in AI systems and the investigation of related concerns, i.e., human, social and ethical concerns, in app reviews.
\subsection{Fairness in AI Systems}
With the increased popularity and adoption of AI-based solutions, the fairness of AI algorithms has attracted the attention of researchers. However, it is evident that ‘fairness’ is a complicated notion with different definitions \cite{narayanan21fairness}. Verna and Rubin \cite{verma2018fairness} argued that while measuring the statistical notion of fairness is relatively easy, more work is required to clarify different fairness definitions and how they are employed in different scenarios. Accordingly, a comparative study of a set of fairness-enhanced ML algorithms by Friedler et al.~\cite{friedler2019comparative} demonstrated that each algorithm could behave differently when it is applied to the same dataset and found that this difference in the fairness behaviour of algorithms arises from issues such as dependency on training data and making significantly different fairness-accuracy trade-offs. Suresh and Guttag~\cite{suresh2021framework} argued that improving the mitigation techniques on fairness-aware ML models requires a thorough understanding of the root causes of the unintended behaviour of these models. Looking into the life cycle of ML models, Suresh and Guttag~\cite{suresh2021framework} highlighted seven sources of harm (e.g., biased, discrimination) in the output of ML models varying from biases in the representation, measurement, and learning phase of a model to biases in aggregation, evaluation of results and deployment of model. Bird et al. \cite{bird2020fairlearn} argued that AI systems have the potential to cause five fairness-related harms, including the reinforcement of existing stereotypes and failure to distribute resources fairly. In a study with 33 AI practitioners, Madaio et al. \cite{madaio2022assessing} found that evaluating the fairness of AI systems comes with three challenges for practitioners: determining the appropriate performance metrics, identifying relevant stakeholders and demographic groups, and collecting datasets. Deshpande and Sharp~\cite{deshpande2022responsible} indicated that tackling the challenge of AI fairness and responsibility requires a better understanding of the stakeholders of AI-based systems. Their study also highlights a broad spectrum of stakeholders that are either impacted by the harm of AI-based systems (e.g., users) or can help to address it (e.g., developers, researchers, tech companies, and legislative agencies). 

\textcolor{black}{Several literature reviews and surveys have been conducted on fairness in AI. Fabris et al. \cite{fabris2022algorithmic} studied over 200 datasets used in fairness research in different domains (e.g., health, natural sciences) to develop standardized and searchable documentation for algorithmic fairness datasets. This effort aims to alleviate the documentation debt present in existing fairness datasets. Hort et al. \cite{hort2023bias} conducted a survey of bias mitigation techniques for ML classifiers. They found that most of the bias mitigation techniques proposed in the literature attempt to mitigate bias during training ML models (i.e., in-processing mitigation techniques), followed by pre-processing mitigation techniques (i.e., those that reduce bias by changing the training dataset). Post-processing is another type of bias mitigation technique that is less explored in the literature. Such techniques are applied after a classifier has been successfully trained. In a review study, Mehrabi et al. \cite{mehrabi2021survey} found that bias and unfairness primarily originate from data and algorithms. Their study also developed a taxonomy of fairness definitions. \textcolor{black}{Another survey by Caton and Haas \cite{caton2024fairness} revealed that most current attempts focus on improving fairness in supervised binary classification. Additionally, they summarized the methods proposed in the literature to measure fairness in ML.}}

Several efforts have been made to measure, avoid, and resolve unfairness in AI-based systems. Madaio et al.~\cite{madaio2020co} developed a fairness checklist through a co-design process with AI practitioners to help organizations develop fair AI-based systems. Bird et al. \cite{bird2020fairlearn} proposed the \textit{Fairlearn} toolkit to help AI practitioners make a trade-off between fairness and model performance. The interactive visualization component of Fairlearn is useful in identifying which group of individuals may be adversely affected by a model. Fairlearn includes several mitigation algorithms and metrics to mitigate unfairness in classifiers. Several researchers have focused on fairness testing and verification (e.g., \cite{biswas2023fairify,galhotra2017fairness,aggarwal2019black,zhang2020white}). Galhotra et al. \cite{galhotra2017fairness} developed a method, \textit{Themis}, to generate test suites for assessing discrimination in software. Biswas and Rajan \cite{biswas2023fairify} have focused on neural network models and proposed \textit{Fairify} to verify individual fairness in such models. Some other studies \cite{li2022training,chakraborty2021bias} have attempted to identify biased labels in datasets.

\subsection{Human, Social, and Ethical Concerns in App Reviews}
The research community has recently attempted to investigate human, social, and ethical aspects of mobile apps, through mining app reviews. For example, some researchers focused on accessibility issues in mobile apps (e.g., \cite{alshayban2020accessibility,alomar2021finding,chen2021accessible,fazzini2022characterizing}). AlOmar et al. \cite{alomar2021finding} built a learning classifier to identify app reviews that include accessibility-related complaints. In the same line of research, Alshayban et al. \cite{alshayban2020accessibility} observed that accessibility issues are common in mobile apps as app developers are not usually trained in accessibility principles. The socio-technical aspect of privacy and security has been investigated in several studies. Nema et al. \cite{nema2022analyzing} proposed a binary classifier to identify app reviews including privacy concerns and a clustering technique to summarize privacy concerns (e.g., “too many permissions”, “too much personal information”). Ebrahimi et al. \cite{ebrahimi2022unsupervised} found that users use domain-specific vocabularies to express their privacy concerns in different application domains (e.g., mental health). They developed an unsupervised summarization technique that can use domain-specific vocabularies to summarize the main privacy concerns in each app category.

Human and ethical values in mobile app development have been the focus of some studies (e.g., \cite{obie2022violation,shams2020society}). Shahin et al. \cite{shahin2023study} studied app reviews to understand the gender-related requirements and concerns discussed by users. To this end, they built a binary classifier to identify reviews that include a gender discussion, followed by a qualitative study. They argued that such gender-related requirements and concerns are essential to building gender-inclusive apps. Khalajzadeh et al. \cite{khalajzadeh2022supporting} constructed a taxonomy that categorizes human-centric issues that users face when working with mobile apps into three high-level groups: “App Usage”, “Inclusiveness”, and “User Reaction”. They then developed several ML/DL classifiers that automatically detect and categorize these three human-centric issues from app reviews. Note that the studies summarized above have either focused on fairness from a technical perspective (learning algorithms and data) or investigated socio-technical aspects such as privacy and honesty other than fairness in app reviews.

Fairness concerns have been only recently investigated in app reviews by Arony et al. \cite{arony2023inclusiveness} as part of the construction of a taxonomy for inclusiveness. The taxonomy was developed based on inclusiveness-related discussions in Twitter, Reddit, and Google Play Store, which includes six major categories: “Fairness”, “Technology”, “Privacy”, “Demography”, “Usability”, and “Other Human Values”. The “Fairness” category has three sub-categories “Terms/Conditions”, “Recommendation”, and “Services”. Arony et al. also leveraged five DL classifiers to automatically distinguish inclusiveness-related discussions from non-inclusiveness-related discussions. \textcolor{black}{Obie et al. \cite{obie2022violation} focused on ‘honesty’ as one of the human values and manually identified 10 types of honesty violations (e.g., ‘delusive subscriptions’, ‘false advertainments’) discussed by users in mobile app reviews. They further developed several ML classifiers to automatically classify user reviews into honesty violation reviews and non-honesty violation reviews. Two types of the identified honesty violations are related to fairness:  ‘unfair cancellation and refund policies’ and  ‘unfair fees’, as the definition of honesty they considered was broad and encompassed fairness, too. While Obie et al. \cite{obie2022violation} and Arony et al. \cite{arony2023inclusiveness} had touched upon some aspects of fairness within their respective scopes of honesty and inclusiveness, their research did not focus on creating a dedicated approach for identifying fairness reviews, automatically classifying types of fairness concerns, or pinpointing the root causes of these concerns.} We argue that there is still a significant gap in understanding the types of fairness concerns raised by users in AI-based mobile app reviews and their potential root causes from the app owners' perspective. In this research, we aim to address this gap with an exhaustive analysis of 108 AI-based apps, specifically analyzing user reviews to extract fairness concerns. Our research not only adds a new dimension to this existing body of research but also brings us a step closer to comprehending fairness from the users' viewpoint.

\section{RESEARCH QUESTIONS AND DATASET}\label{sec:rqdataset}

\textcolor{black}{This paper aims to deeply analyze fairness in AI-based app reviews.} This will be achieved by answering the following overarching research questions (RQs).

\begin{tcolorbox}[arc=0mm,width=\columnwidth,
                  top=0mm,left=0mm,  right=0mm, bottom=0mm,
                  boxrule=.75pt]
\textcolor{black}{\textbf{RQ1}: Can we effectively identify fairness reviews in AI-based app reviews?}
\end{tcolorbox}

\textcolor{black}{\textbf{Motivation}. The users share a wide range of information and opinions on app stores, including socio-technical issues they experience when using mobile apps~\cite{wang2022your}. The richness of feedback shared by users has motivated researchers and app providers to mine app reviews, leading to extracting actionable information for app owners and developers. We argue that user reviews of AI-based mobile apps can be a rich resource for uncovering fairness concerns from the users' perspective. However, manually identifying fairness reviews in millions of app reviews is challenging. While automated techniques, including ML and DL models, have been proposed in recent years for mining app reviews (e.g.,~\cite{wang2022your,nema2022analyzing,obie2022violation,khalajzadeh2022supporting}), none of these automated techniques has focused on fairness concerns. Furthermore, applying general-purpose automated techniques may misidentify and misclassify fairness concerns. To help close this gap, we plan to experiment with several models to automatically distinguish fairness reviews from non-fairness reviews.}

\begin{tcolorbox}[arc=0mm,width=\columnwidth,
                  top=0mm,left=0mm,  right=0mm, bottom=0mm,
                  boxrule=.75pt]
\textbf{RQ2}: What types of fairness concerns are raised by users in AI-based app reviews?
\end{tcolorbox}

\textcolor{black}{\textbf{Motivation}. Distinguishing fairness reviews from non-fairness reviews is the first step to understanding fairness reviews. Still, it cannot provide a fine-grained view of different types of fairness concerns raised by various users of AI-based apps. Hence, this RQ aims to use a clustering technique to summarize the fairness concerns raised in the fairness reviews identified from RQ1.}

\begin{tcolorbox}[arc=0mm,width=\columnwidth,
                  top=0mm,left=0mm,  right=0mm, bottom=0mm,
                  boxrule=.75pt]
\textbf{RQ3}: What are the root causes of fairness concerns in apps, based on the justification by app owners?
\end{tcolorbox}

\textbf{Motivation}. The Google Play Store provides a mechanism that allows app owners to respond to the posted user reviews. App owners often use this opportunity for various purposes such as `appreciating users', `seeking further information from users', `providing solutions for the issues raised by users', and `justifying the issues raised by users'~\cite{hassan2018studying,chen2021should}. This RQ aims to identify the root causes reported by app owners to justify fairness concerns. To this end, we qualitatively analyze app owners' responses to fairness concerns.

\subsection{Dataset Curation}\label{sec:dataset}
\begin{figure*}
    \centering
    \includegraphics[width=.99\linewidth]{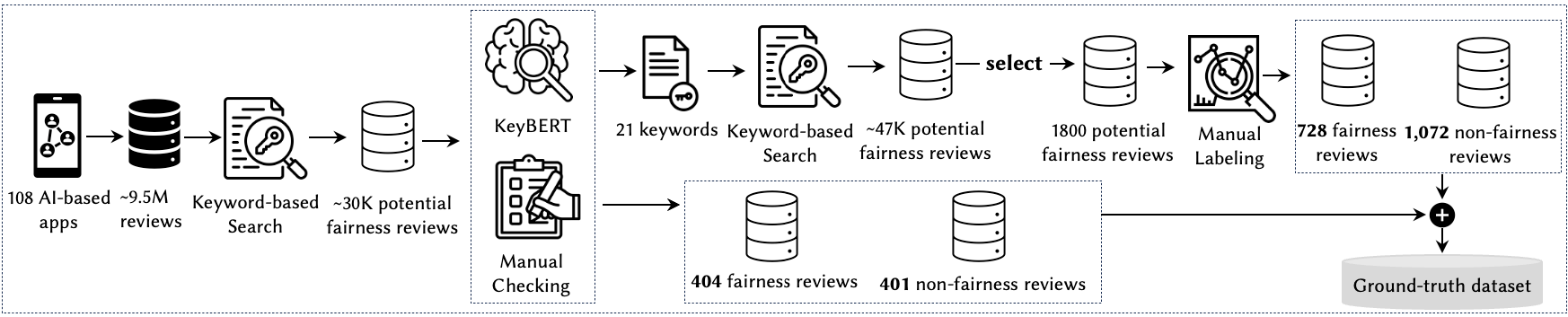}
    \caption{Overview of Dataset Curation}   
    \label{fig:dataset}
    \vspace*{-1em}
\end{figure*}

There is no current dataset (to the best of our knowledge) with fairness and non-fairness app reviews. We curated a dataset to evaluate our learning models and clustering techniques. Fig.~\ref{fig:dataset} illustrates our dataset curation process, discussed next. 

\input{Tables/AIapps}
\subsubsection{Identification of AI-based apps} Several studies have developed approaches to identify AI-based apps in Google Play Store. To have a relatively comprehensive list of AI-based apps, we used the existing datasets of AI-based apps~\cite{xu2019first,li2022ai}. Xu et al. \cite{xu2019first} identified 211 offline-mode DL Android apps in Google Play Store by examining app installation packages. \textcolor{black}{They defined DL Android apps as mobile apps using one of the 16 popular DL frameworks such as TensorFlow Lite \cite{TensorFlowlite}, Caffe2 \cite{caffe2}, etc.} As DL is only one sub-category of AI techniques, Li et al. \cite{li2022ai} extended this to develop an approach called AI Discriminator that works based on app dissection and keyword matching and identifies apps that use any AI frameworks. Li et al. collected 56,682 AI-based apps in the AndroZoo repository (i.e., a collection of Android apps \cite{allix2016androzoo}). \textcolor{black}{They defined AI-based apps as apps that use any of 33 AI frameworks categorized into classic ML-based, DL-based, and AI service-based frameworks (e.g., Google AI \cite{GoogleAI}).} The two datasets had 56,893 AI-based apps in total. We removed duplicates from these two datasets and removed apps with less than 10K reviews, i.e., only included `popular' apps with sufficient user feedback. This process led to 108 AI-based apps. These apps come from 23 distinct categories (e.g., social, communication, and productivity). \textcolor{black}{Table \ref{tbl:AIApps} shows the number of AI-based apps in each category.} Our replication package \cite{replicationpackage} provides a demographic of these apps.

\subsubsection{Collection of potential fairness reviews} \label{sec:colfairreviews} The next step was to collect user reviews from the 108 AI-based apps. Google Play Store can present the reviews of an app in different ways, such as in the order of they were written or by relevance. To filter out short reviews (e.g., one or two words long) or not very informative reviews, we decided to use the most relevant reviews sorted by Google Play Store. Given that the 108 AI-based apps had different numbers of reviews and our aim was to build as large, diverse, and balanced dataset as possible, we collected the top 500K reviews from each app with more than 500K reviews using a Google Play Crawler~\cite{googleplaycrawler}. For the remaining apps, we collected all of their reviews. This process resulted in 17,968,298 reviews. Next, we removed reviews with less than 4 words and non-English reviews~\cite{dkabrowski2022analysing}. In the end, we were left with 9,475,506 reviews that potentially covered several topics such as feature requests, bug reports, non-technical concerns, etc. We also expected that fairness concerns are less discussed than other issues such as accessibility issues in reviews. Hence, we aimed to find reviews that were highly likely to include fairness concerns (we call such reviews `\textbf{potential fairness reviews}'). To this end, we iteratively built a fairness keywords-set using manual and automated steps. The rationale behind applying both manual and automated steps iteratively was that we wanted to reduce the chance of having a large number of uncontrolled false positives and to minimize the chance of excluding relevant keywords. The following steps show how potential fairness reviews were identified.

\underline{1. Search with an initial keywords set.} While fairness is a confusing term and can have different meanings in different contexts~\cite{ekstrand2022fairness}, our observation of exploring different literature reviews (e.g., \cite{xivuri2021systematic,pessach2022review,mitchell2021algorithmic,mehrabi2021survey,chen2022fairness}) and grey literature (e.g., white papers issued by governments and institutions~\cite{dawson2019artificial,anderson1992acm,MSResponsible}) on fairness shows that `fair', `discrimination', and `bias' are three commonly used keywords to refer to fairness. We performed a search using these three keywords on our set of 9,475,506 reviews, which resulted in a set of 30,019 `potential fairness reviews'.

\underline{2. Using KeyBERT.} ‘Fair’, ‘discrimination’, and ‘bias’ are not the only keywords that users use to discuss a fairness concern. Hence, we used KeyBERT \cite{grootendorst2020keybert}, a BERT-based keyword/phrase extraction technique, to extract more relevant fairness-related keywords and phrases, thereby increasing the representativeness and comprehensiveness of the final dataset. There are many embedding models that can be used in KeyBERT. We decided to use the \hbox{\textsf{all-mpnet-base-v2}} embedding model as it is known to work well with English texts~\cite{pretrainedmodels}. \textcolor{black}{We conducted several pilot experiments to understand if KeyBERT should be applied to all 9M reviews or the 30,019 potential fairness reviews. Our experiments revealed that applying KeyBERT to all 9M reviews led to many irrelevant suggested keywords and their corresponding reviews. Hence, KeyBERT with the initial keywords (`fair', `discrimination', and `bias') as seed input was applied to the 30,019 potential fairness reviews to extract uni-gram, two-gram, and three-gram keywords. This approach produced less noise while identifying the most relevant keywords and fairness reviews.} Given that the 30,019 reviews cover different topics, adding the seed keywords guided KeyBERT to look for keywords that are most similar and best describe fairness concerns in reviews.

KeyBERT suggested 210 keywords, including three initial keywords. In the next step, the second and third authors manually checked these keywords. They removed duplicates, the suggested 2-gram or 3-gram keywords (e.g., ``fair policy'') that we had their uni-gram keywords (e.g., ``fair''), and keywords that were rarely found in reviews (e.g., less than 0.001\% of 9,475,506 reviews). This step resulted in reducing the number of keywords from 210 to 42 keywords. To make sure these 42 keywords are relevant and do not lead to unmanageable false positives, we next randomly selected 20 reviews per keyword matching, i.e., $42*20=840$ reviews. We removed duplicates in 840 reviews and reached 805 reviews. It was because some keywords appeared in more than one review. \textcolor{black}{The second and third authors independently labeled these 805 reviews. Given the absence of a universally accepted definition of fairness, the second and third authors were asked to review various definitions of fairness from different domains. These definitions were collected from both peer-reviewed literature \cite{bird2020fairlearn,mehrabi2021survey,pessach2022review,d2024fair, liu2024faircompass} and gray literature \cite{dawson2019artificial,anderson1992acm,MSResponsible,IEEE_Ethics_Standard}. For example, Pessach and Shmueli \cite{pessach2022review} argued that the legal domain defines discrimination in two ways: direct discrimination, which occurs when an individual is intentionally treated differently based on his/her membership in a protected class, and indirect discrimination, which occurs when members of a protected class are negatively affected more than others even if by a seemingly neutral policy. This ‘definition review’ process was intended to guide their labeling decisions on whether a particular review should be classified as a fairness review. The inter-rater agreement between the second and third authors was measured using the Cohen's kappa coefficient \cite{mchugh2012interrater}, which yielded a value of 0.69. This indicates a substantial level of agreement. The two authors had 126 disagreements in this step, which were resolved using the negotiated agreement approach~\cite{campbell2013coding}. The main reason behind the disagreements between the two authors could be attributed to the fact that they have different understandings of fairness and the definitions of fairness provided to them. At the end of this step, the 805 reviews were classified into 404 fairness reviews and 401 non-fairness reviews (see Fig.~\ref{fig:dataset})}. 

In this process of manual classification, we figured that 21 of these 42 keywords were irrelevant or redundant. This was determined by either (i) none of the matched reviews using these keywords was labeled as fairness review, or (ii) the matched reviews of another keyword subsumed all reviews matched by this keyword. For example, one of the irrelevant keywords was ``suspend''. A review that matched this keyword was \textit{``My drive got suspended because of a DMCA (Digital Millennium Copyright Act) claim and I wasn't active. I don't have access to my personal data.''} This review was labeled non-fairness review because it does not discuss anything about fairness. Next, we searched the remaining 21 keywords on 9,475,506 reviews and got 46,700 potential fairness reviews. \textcolor{black}{Inspired by the collected definitions of fairness and our observations in this step, we defined a fairness review as follows: a review in which users express that they are treated differently, whether intentionally or unintentionally, due to their inherent, acquired, or context-specific attributes (e.g., age, language, location, gender) by apps or the decisions and policies of the app owners. Additionally, a fairness review can reflect users' perceptions that the outcomes of apps, or the decisions and policies of the app owners, are inconsistent, non-transparent, and unreasonable. In such fairness reviews, users do not often explicitly compare themselves with others based on inherent, acquired, or context-specific attributes. Instead, they typically complain about outcomes, decisions, and policies that they perceive as inconsistent, non-transparent, and unreasonable (e.g., \textit{``I used to love this app, but since last month, for no reason, my account has been getting restricted again and again every 24 hours. I am blocked. [the app name] should do something about it; that's unfair for users like me who have never used any third-party app, bot, or spam and still suffer the restrictions and shadow ban.''}).}

\subsubsection{Manual labeling} \label{manuallabel}
The previous step resulted in 404 fairness reviews, 401 non-fairness reviews, and 46,700 potential fairness reviews. The 46,700 potential fairness reviews come from 108 apps. Note that these 108 apps possessed varied numbers of potential fairness reviews. Our aim was to build a reasonable and roughly balanced and diverse dataset of potential fairness reviews that reflect users' opinions of different apps from various categories. To this end, we decided to randomly choose 45 potential fairness reviews from the apps that had at least 45 potential fairness reviews. This step resulted in 1,800 potential fairness reviews from 40 apps. This number of reviews (1,800 reviews) is statistically representative as it well exceeds the number needed for a statistical sample with a 99\% confidence level and a 3\% margin of error~\cite{israel1992determining, ahmad2017determining}. 

\textcolor{black}{The labeling process of 1,800 potential reviews was conducted in three rounds and included five coders with extensive experiences in the human and social aspects of software engineering. This helped avoid fatigue, gradually reach a common understanding, and better manage disagreements. In each round, $\approx$300 reviews were independently labeled by two coders. The five coders included the two coders who were involved in the previous step and three new coders (i.e., the second, fourth, and fifth authors). While the definition of fairness reviews was given to all coders, including new coders, to guide labeling their allocated reviews, they also accessed the various fairness definitions collected in the previous step. The inter-rater agreement between the coders was quantified using the Cohen's kappa coefficient \cite{mchugh2012interrater}, resulting in a coefficient of 0.75, which shows a substantial agreement. This level of agreement indicates high reliability in the coding process.} In total, 227 disagreements were found between the coders. In case of any disagreement between two coders, a third coder was asked to read the review and label it. Hence, the final label was determined based on majority opinion. Our labeling process led to 728 fairness and 1,072 non-fairness reviews, which with 404 fairness and 401 non-fairness reviews in Section~\ref{sec:colfairreviews} constitute a dataset (ground-truth) of 1,132 fairness and 1,473 non-fairness reviews. Those potential fairness reviews labeled as non-fairness mainly reported a bug or did not discuss any concrete fairness concerns (e.g., \textit{``My video is not saving in my Gallery and it's stopping at 80\% then saved in drafts only. This is not so fair''} and \textit{``I don't think it's fair}'').

\section{Fairness Reviews Identification (RQ1)}\label{sec:fairnessclassifier}

\subsection{Approach} This section discusses the development of our model that accurately distinguishes fairness reviews from non-fairness reviews. \textcolor{black}{For this purpose, we decided to use a wide range of classification models, including classical ML models, state-of-the-art (SOTA) pre-trained language models (LMs), and large language models (LLMs). We opted to experiment with various models because (1) it was unknown to us from the beginning which model could produce the best result and the feature-based ML techniques have been known to outperform SOTA pre-trained LMs \cite{ezzini2022automated}; (2) although the literature has shown the promising performance of ML and DL models in text classification tasks, their performance could vary on different text-based datasets \cite{aggarwal2012survey,caruana2006empirical,abdalkareem2020machine}; and (3) classifiers use different strategies to handle overfitting and explainability, and have different execution times~\cite{kotsiantis2006machine, abdalkareem2020machine}.}

\subsubsection{Dataset} The dataset constructed in Section \ref{sec:dataset} was used to train and test our learning models. The dataset contains 1,132 fairness reviews and 1,473 non-fairness reviews. Following the NLP best practices applied to app reviews \cite{dkabrowski2022analysing}, we cleaned the dataset by removing emojis, numbers, punctuation, and one-letter characters. 

\subsubsection{\textcolor{black}{Models}}\label{subsec:models}
In total, we used ten models. Here, we describe six of them that performed better than the other four, due to space constraints. Previous studies \cite{aniche2020effectiveness, fu2016tuning, menzies2012special} indicated that the parameters of these prediction models could be fine-tuned to improve their performance. Following the recommendations in \cite{bao2019large, tantithamthavorn2018impact}, we checked several scenarios and values for the hyperparameters of each prediction model to find the best configuration for each prediction model. Specifically, we discuss how we implemented and fine-tuned them to achieve their best performance. The implementation and configuration of the rest of the models can be found in our replication package~\cite{replicationpackage}. 

\textcolor{black}{\textbf{Traditional ML Models.}
In the traditional ML models, we used three common text feature extraction techniques: Term Frequency–Inverse Document Frequency (TF-IDF), Universal Sentence Encoder (USE), and Word2Vec. Initially, we briefly describe these feature extraction methods and the configuration we used to generate feature vectors from the reviews in our dataset. Then, we explain the fine-tuning of the traditional ML models.}

\textit{TF-IDF} calculates the significance of a word in a text corpus~\cite{ramos2003using}. We configured TF-IDF to generate a feature union by capturing 4-, 5-, and 6-gram characters from the reviews. We also set the feature union to the top 50,000 features~\cite{vogel2020fake}. \textit{Word2Vec} is a word embedding technique that captures the semantic and syntactic relationships between words in a large corpus of text~\cite{mikolov2013efficient}. 
We used the pre-trained 300-dimensional Google News Word2Vec embedding model containing 3M words and phrases, to produce a vector for each word in reviews of our dataset~\cite{mikolov2013distributed}. \textit{USE} works at the sentence level and utilizes pre-trained sentence embedding models to produce sentence vectors~\cite{cer2018universal}. 
We used the DAN encoder to encode a review into a 512-dimensional vector, as DAN encoders are better suited for large-scale tasks and are efficient regarding computational resources. As app reviews analysis, can be based on many reviews, we decided to go for DAN-based USE.

Research has shown that using more than one feature extraction technique may help to yield better results in ML models \cite{shahin2023study}. Hence, we also experimented with the various combinations of these three techniques to understand if their combinations enhance the performance of the ML models.

\textit{LR (Logistic Regression)} is a widely used model for binary classification tasks, which estimates the probability of a positive class given input features~\cite{dreiseitl2002logistic}. Considering the size of our dataset, we regularized the solver parameter to \hbox{\textsf{``liblinear''}} \cite{scikitLR}. Also, the penalty was set to L2, which could shrink the coefficients of the features and reduce overfitting. 

\textit{SVM (Support Vector Machine)} uses mathematical principles to find the best way to classify data into different classes \cite{vapnik1999nature}. In this study, we set the kernel parameter of this model to linear to compute the similarity of two input vectors. 

\textit{XGBoost (XGB)} (Extreme Gradient Boosting) implements a regularized version of gradient boosting for classification problems \cite{chen2015xgboost}. 
To have a maximum performance of XGBoost, we set the number of gradient boosted trees to 200 and the maximum depth of each tree to 5. 

\textcolor{black}{\textbf{Pre-trained Language Models.} In this study, we used two widely used pre-trained models to detect fairness reviews automatically.}

\textit{BERT (Bidirectional Encoder Representations from Transformers)} is a neural network architecture that can learn vector-space representations of natural language for various downstream tasks \cite{devlin2018bert}. We encoded the reviews in our dataset to a sequence of vectors using the Hugging Face BERT tokenizer that implements the WordPiece tokenization algorithm \cite{hugfacebert, hugfacetoken} and \hbox{\textsf{``bert-base-cased''}}, a pre-trained model with 110 million parameters \cite{huggingfacetrainedmodels}.

\textit{RoBERTa (Robustly Optimized BERT Pretraining Approach)} is a modified version of BERT that enhances performance on various NLP tasks~\cite{liu2019roberta}. As an input to RoBERTa, the reviews in our dataset were transformed to a sequence of vectors using the Hugging Face RoBERTa tokenizer that uses the byte-level Byte-Pair Encoding (BPE) algorithm and \hbox{\textsf{``roberta-base''}}, a pre-trained model with 125 million parameters~\cite{huggingfacetrainedmodels}. We used linear transformation in both BERT and RoBERTa to map the hidden states to the output vocabulary size. Then, a combination of the sigmoid layer and binary cross-entropy loss was used as the last layer in both models to categorize a review into a fairness review or a non-fairness review. We also used the Adam algorithm \cite{kingma2014adam} with a learning rate of 2e-5 to optimize the models and trained their networks for 10 epochs with a batch size of 15.

\textcolor{black}{\textbf{Large Language Models (LLMs).}} Finally, we used \textit{GPT-3 (Generative Pre-training Transformer)} developed by OpenAI\footnote{https://openai.com} to identify fairness reviews. \textit{GPT-3} is a large language model trained on a huge amount of text data that can produce human-like texts for various tasks \cite{brown2020language}. Its performance can be further improved with human feedback \cite{brown2020language}. GPT-3 uses BPE (Byte-Pair Encoding) as a tokenizer to map a given text into a sequence of integers that represent the text. This process relies on \hbox{\textsf{``text-davinci-003''}}, a pre-trained model with 175 billion parameters built on top of the InstructGPT model \cite{brown2020language}.

\input{Tables/metrics}


\subsubsection{Evaluation Metrics}
\textcolor{black}{Similar to other studies \cite{yang2022survey, abdalkareem2020machine,viviani2019locating}, we used precision, recall, F1-score, accuracy, and AUC to show the performance of the models. As shown in Table \ref{tbl:Metrics}, these five common metrics are calculated based on False Positive (FP), True Positive (TP), True Negative (TN), and False Negative (FN). Note that AUC=1.0 signifies a perfect model (i.e., able to perfectly discriminate between the two classes), while an AUC=0.5 suggests that the model is no better than random guessing; a higher AUC value closest to 1.0 is desired \cite{berrar2019performance}.}

\textcolor{black}{We used the 10-fold cross-validation method to evaluate the predictive capabilities of the traditional ML models and pre-trained language models \cite{efron1983estimating}.} The 10-fold cross-validation method splits our dataset (ground-truth) into ten parts (i.e., fold), which are used for training and testing the models ten times. In each iteration, nine parts (90\% of the dataset) are used as training data and one part for evaluating the models. We calculated the performance of GPT-3 by comparing the labels (fairness or not-fairness) generated by GPT-3 and the labels of the reviews in the ground-truth dataset. We used the GPT-3 API and simply asked GPT-3 to determine if reviews in the dataset are about fairness concerns or not (i.e., by prompting \textit{``Tell me if the following text is related to a fairness concern or not. Just say Yes or No''}).

\input{Tables/binaryclassifications}

\subsection{Results}

\subsubsection{Performance of the models}\label{sec:binaryresults} Table~\ref{tbl:MLResults} shows the performance of the models (described in Section \ref{subsec:models}) in detecting fairness reviews. Given that we used 10-fold cross-validation for ML and pre-trained models, the score in each row, except the last row (i.e., GPT-3), is the average over the ten folds. Among the models, RoBERTa achieves the highest F1-score (0.82), followed by BERT with an F1-score of 0.80. Considering the accuracy metric, SVM with the combination of TF-IDF and USE as the feature extraction technique reaches the highest accuracy of 0.83, meaning that 83\% of predictions (either fairness or non-fairness) are correct. As shown in Table \ref{tbl:MLResults}, several models such as RoBERTa, Logistic Regression, BERT, and XGBoost reach an accuracy of above 0.80. We see while GPT-3 achieves the highest recall of 0.81, it has low precision of 0.63, meaning that it produces low false negatives and relatively high false positives (more non-fairness reviews are labeled as fairness). RoBERTa with a recall of 0.78 and BERT with a recall of 0.77 are the second-best and third-best models, respectively, from the recall perspective.

Table~\ref{tbl:MLResults} indicates that Logistic Regression with TF-IDF has higher precision (0.88) than all other models but quite low recall (0.49). Using the combination of three feature extraction techniques (TF-IDF, USE, and Word2Vec) leads to increasing the recall of Logistic Regression from 0.49 to 0.70 and having a precision of 0.85. RoBERTa has also a high precision of 0.85 and a reasonable recall of 0.78. Considering the AUC scores, Logistic Regression with TF-IDF, USE, and Word2Vec performs better than other models.

Given that our ultimate goal is to identify different types of fairness concerns from the user perspective, we decided to select Logistic Regression with TF-IDF, USE, and Word2Vec as our best model (see Table \ref{tbl:MLResults}). It is because (1) on average, this model has the highest AUC, high precision and good F1-score. (2) Logistic Regression with TF-IDF, USE, and Word2Vec reaches a precision of 0.94, AUC of 0.93, and recall of 0.77 in the seventh fold. In other words, inspired by the methodology used in \cite{nema2022analyzing}, we prioritized AUC and precision over recall. We used this model that reaches \textbf{a precision of 0.94} in the seventh fold for the rest of our work (Section \ref{sec:distribution}, Section \ref{sec:patterns}, and Section \ref{sec:cluster}). Particularly, this model could minimize the number of noises in our clusters (see Section \ref{sec:cluster}).  

\input{Tables/unseentable}

\begin{figure}
    \centering
    \includegraphics[width=0.5\linewidth]{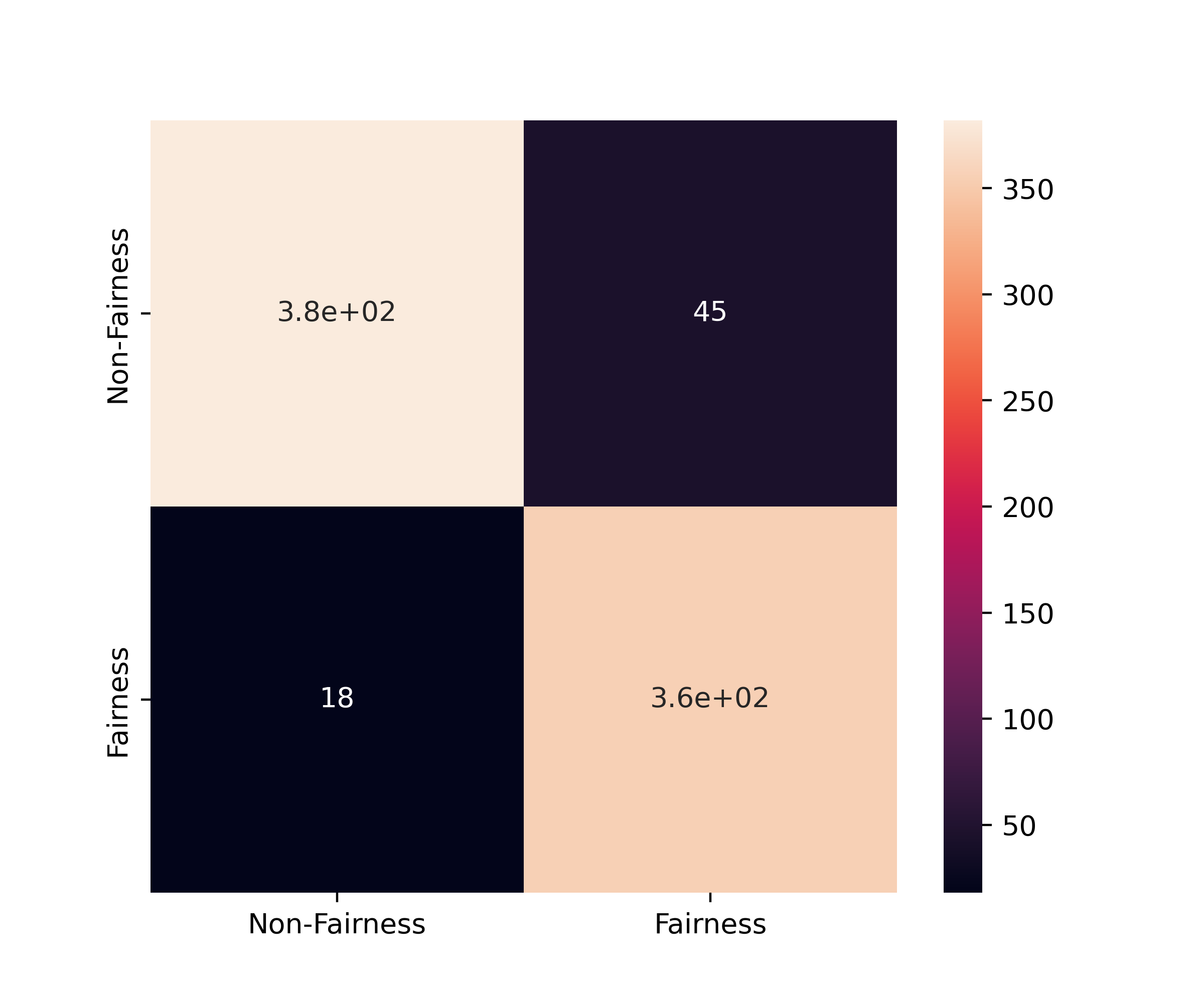}
    \caption{\textcolor{black}{Confusion matrix of our best performing model for all reviews}}
    \label{fig:unseendata}
    \vspace*{-0.6em}
\end{figure}

\subsubsection{\textcolor{black}{Performance of the best-performing model on all reviews}} \label{sec:unseen}
\begin{figure*}
    \centering
    \includegraphics[width=0.85\linewidth]{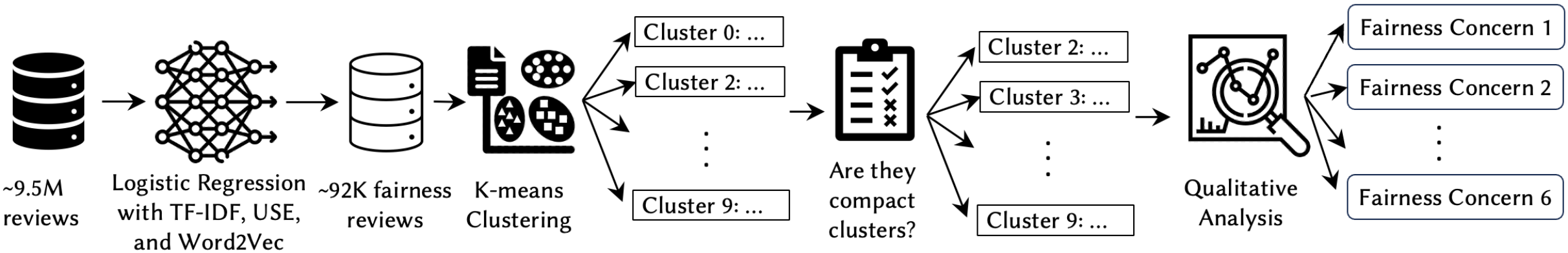}
    \caption{Applying the best model and K-means}
    \label{fig:binarykmeans}
    \vspace*{-0.6em}
\end{figure*}

\textcolor{black}{As discussed in Section \ref{sec:colfairreviews}, we collected 9,475,506 reviews from 23 distinct app categories. As shown in Fig. \ref{fig:binarykmeans}, we applied our best model (Logistic Regression with TF-IDF, USE, and Word2Vec) on this number of reviews. The best model classified the 9,475,506 reviews into 91,974 fairness reviews and 9,383,532 non-fairness reviews. To evaluate the performance of the best model on all reviews, we randomly selected 400 reviews from 91,974 non-fairness reviews and 400 reviews from 91,974 fairness reviews with a confidence level of 95\% and a margin of error of 5\% \cite{israel1992determining, ahmad2017determining}. The first and third authors undertook the task of independently labeling these 800 reviews. The inter-rater agreement between them was a Cohen's kappa coefficient of 0.92, indicating perfect agreement. They had 30 disagreements, which were resolved through the negotiated agreement approach~\cite{campbell2013coding}. Table \ref{unseentab} shows that applying the best model on all reviews yielded good results in all metrics: a precision of 88.75\%, a recall of 95.17\%, an F1 of 91.85\%, an ACC of 92.12\%, an AUC of 92.32\%. Fig. \ref{fig:unseendata} also shows the confusion matrix of the best model for all reviews.}

\subsubsection{Distribution of fairness reviews across app categories} \label{sec:distribution} \textcolor{black}{We also wanted to know the distribution of fairness reviews in different app categories. Section \ref{sec:unseen} shows that 91,974 reviews (0.97\%) out of all reviews (9,475,506) were classified as fairness reviews. Table \ref{tbl:fairnessincategories} indicates the number of reviews and the percentage of fairness reviews in each app category.} As shown in Table \ref{tbl:fairnessincategories}, the `communication' category has the highest percentage of fairness reviews (1.76\%), followed by the `social' category (1.64\%) and `entertainment' category (1.45\%). We also found that app users in the `personalization' and `auto \& vehicles' categories rarely discuss fairness issues in reviews (i.e., less than 0.1\% of their reviews are about fairness). 

\subsubsection{Patterns in the extracted fairness reviews} \label{sec:patterns} 
Our analysis shows that the models, particularly the best-performing model, can identify reviews that do not include any of the 21 fairness-related keywords collected in Section \ref{sec:dataset}. For instance, this model correctly classified the following reviews as fairness reviews: \textit{``I love the app but I need a dark mode. I'm on Android and I know that iOS has a dark mode option available. Dark mode would make this 5 stars.''} and \textit{``[The app is] not good in the language. [It] does not support the Arabic language''}. In the first review, the user seemingly criticizes the app provider as it does not provide the same dark mode feature that is available for iOS users to Android users. The second review complains about the app as it does not support the Arabic language. In both examples, such unfair behaviours may lead to the exclusion of or discrimination towards some users. We also found some patterns in the reviews misclassified by our model that can be investigated in future to enhance the model. Some of these reviews include one or more of the 21 fairness-related keywords but they do not raise any concrete or specific fairness concerns or discuss root causes of raised fairness concerns. For example, the following review \textit{``Easy to use and alerts are prompt! [I] just wish I could get my hands on the tech grey. For people [who] have been loyal customers literally their entire lives, I think it's unfair not to be able to purchase a product they really want but can't due to it being sold out.''} criticizes a product produced by a company and does not point out any concrete fairness concern regarding the app. We labeled such reviews as non-fairness reviews in our ground-truth dataset. However, our model sometimes classifies such reviews as a fairness review (a false positive review). It is possibly because the model makes the choice based on a fairness-related keyword (e.g., ``unfair'' in the above review) and the negative sentiment of the review. This may suggest that our model needs a more diverse and larger dataset for training and testing.

\input{Tables/fairnessincategories}

\begin{tcolorbox}[arc=0mm,width=1.02\columnwidth,
                  top=0mm,left=0mm,  right=0mm, bottom=0mm,
                  boxrule=.75pt]
\textbf{Key Finding 1}: Our experiments show that Logistic Regression, combining TF-IDF, USE, and Word2Vec as the feature extraction technique, is the best-performing model to detect fairness reviews and reaches the highest precision of 94\%. 
\end{tcolorbox}

\section{Fairness Clustering and Summarization (RQ2)} \label{sec:cluster}

As discussed in Section~\ref{sec:fairnessclassifier}, the models can accurately identify fairness reviews from a large number of reviews. However, they do not provide a fine-grained view of different types of fairness concerns raised by various users of AI-based apps. Hence, we aimed to cluster and summarize fairness reviews, as discussed next. 
\subsection{Approach} Here we discuss our clustering and summarization approach to identify different fairness concerns in app reviews.

\subsubsection{Dataset} As discussed in Section \ref{sec:binaryresults}, Logistic Regression was selected as the best-performing model because it had the highest precision and AUC. Once we applied it on 9,475,506 reviews, we got 91,974 fairness reviews (see Fig. \ref{fig:binarykmeans}). We used these 91,974 reviews as a dataset for the clustering and summarization approach. 

\subsubsection{K-means Clustering} \label{sec:kmeans} Many clustering techniques could be used for clustering fairness reviews \cite{saxena2017review}. In this study, we selected K-means clustering because it is one of the simplest, widely used and benchmarked clustering approaches~\cite{macqueen1967some,arora2016automated} and has recently been successfully used by the software engineering researchers to cluster app reviews (e.g., identifying privacy themes in reviews) \cite{nema2022analyzing, wang2022your}. K-means partitions a given set of data points or the reviews in our context into K clusters, where $k$ is an algorithm parameter. K-means attempts to assign each review to one of the $k$ clusters by maximizing the similarity between the individual reviews in each cluster and the center of that cluster. We represent each review for K-means clustering using a 512-dimensional embedding from the USE model (see Section~\ref{subsec:models}).
As mentioned above, one needs to know $k$ apriori in K-means. In our case, we had no idea how many distinct fairness concerns ($k$ fairness clusters) users may discuss in app reviews. To this end, several metrics have been proposed to address this common challenge in clustering tasks~\cite{saxena2017review}. 
We decided to use the \textit{summarization metric} suggested by Nema et al.~\cite{nema2022analyzing} to identify the suitable value of \textit{k} in our work. Nema et al. used the \textit{summarization metric} to extract and summarize privacy concerns in app reviews. The summarization metric is designed to find \textit{k} focusing on minimizing the influence of the following two issues in data, i.e., (i) a data point can appear in multiple clusters produced by K-means and (ii) a data point discusses several topics or is an outlier. We chose the summarization metric because we had such issues in our fairness reviews. For example, the following fairness review discusses at least two distinct `fairness concerns' (i.e., mixed-concern fairness reviews).

 ``\textit{I'm a [the app name] Premium subscriber, ... \textbf{too many ads, weak and unfair policies}, no support to the creators who made the platform what it is. \textbf{Unfair algorithms, bad recommendations} and we can no longer comment what we want or see how many dislikes there are... How can we see what the perception of the public is?}''

In the first one, the user deems the policy adopted by the app to include ads in the videos as unfair for premium subscribers. In the second one, the user thinks that the employed algorithms do not fairly recommend videos, based on the user's preferences.

\begin{figure}
    \centering
    \includegraphics[width=0.80\linewidth]{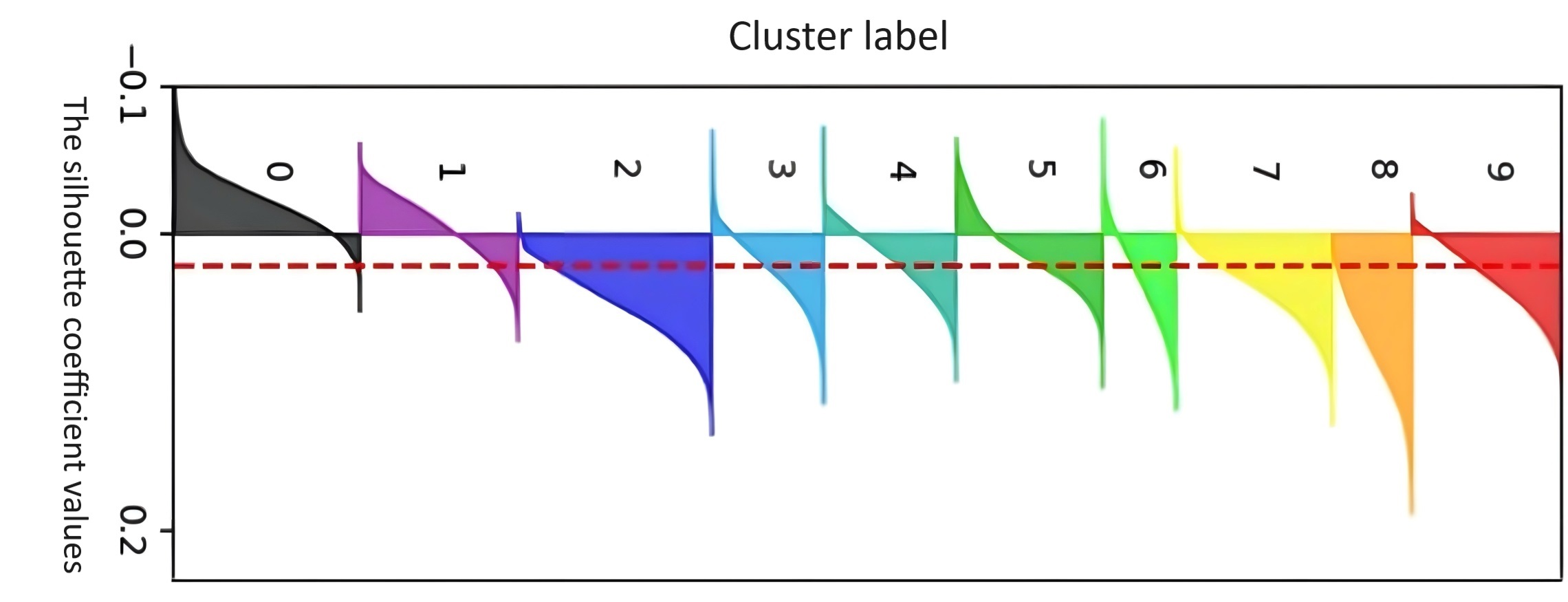}
    \caption{The silhouette scores for each cluster of k=10}
    \label{fig:silhouettescore}
\end{figure}

In addition to the summarization metric, we used the silhouette score to determine the quality of clusters~\cite{calinski1974dendrite}. The silhouette score is a measure of how similar a data point is to the other data points in its cluster compared to how similar it is to the data points in other clusters. \textcolor{black}{A low silhouette score of a review implies that the review could be equally well assigned to another cluster.} The silhouette score thus indicates whether a given cluster is well-formed or not. We executed K-means with different $k$ ranging from 2 to 10. \textcolor{black}{We did not go beyond 10 as our observations in the labeling process (see Section \ref{sec:dataset}) provided clues that there should not be more than 10 dominant fairness concerns in user reviews. However, as discussed in \cite{nema2022analyzing}, a larger $k$ could have helped us identify the long-tail of fairness concerns.} 
For each $k$, K-means was run 100 times with different centroid seeds. We also set the maximum number of iterations for a single run to 100. The final result of K-means is the best output out of 100 runs in terms of inertia, which is the sum of squared distances of samples to their closest cluster center. Our experiments showed that the highest value for the summarization metric is achieved when \textit{k=10} (Fig.~\ref{fig:binarykmeans}). Fig.~\ref{fig:silhouettescore} presents the distribution of the silhouette scores of reviews in each of the 10 clusters (\textit{k=10}). The dotted red line shows the average silhouette score for \textit{k=10}. As shown in Fig. \ref{fig:silhouettescore}, at least 40\% of the reviews in clusters 2 to 9 have a silhouette score more than the average (the dotted red line). Clusters 2 to 9 are considered \textbf{compact clusters}. \textcolor{black}{Clusters 0 and 1 were excluded from further analysis because they did not demonstrate a high degree of cohesion, as more than 60\% of the reviews in these clusters had a silhouette score below the average.}

To understand the main (dominant) fairness concern discussed in each of the eight compact clusters, we descendingly sorted the reviews in each compact cluster based on their silhouette scores. The first author manually analyzed the 30 reviews with the highest silhouette scores in each cluster ($8*30=240$ reviews in total) and suggested a topic for each cluster. Then, three other authors rechecked all these 240 reviews and the proposed topics. The first author and the rest of the authors held several meetings to discuss the topics, refine them, and solve any disagreements. \textcolor{black}{Although we had eight compact clusters, as shown in Fig. \ref{fig:binarykmeans}, the qualitative analysis process led to the identification of \textbf{six distinct, dominant fairness concerns} that are discussed next in Section~\ref{sec:sumresults}. This was achieved by merging the topics emerging from some clusters that exhibited overlapping concerns. This approach was taken to provide a more concise and distinct set of fairness concerns. For example, we identified two clusters with approximately overlapping topics. One highlighted user concerns about `disparities in feature quality and service access across platforms and devices', and another included complaints about `being deprived of some features and services in platforms and devices'. Given the closeness of these concerns, we merged them into a single fairness concern called `receiving different quality of features and services in different platforms and devices'. This new fairness concern encapsulates the essence of both initial fairness concerns. Merging fairness concerns was done thoughtfully to ensure that the distinct nature of each concern was preserved while avoiding redundancy.}

\subsection{Results}\label{sec:sumresults} This section summarizes the top six fairness concerns (fairness topics) raised by users in AI-based app reviews. The representative review examples in each fairness concern come from the top thirty reviews in each cluster (see Section \ref{sec:kmeans}). We deliberately chose examples from different apps and app categories to show that fairness concerns are not limited to a certain type of app.

\textbf{Fairness Concern 1: Receiving different quality of features and services in different platforms and devices.} Apps are usually available on different platforms and devices. Our analysis found that many users complain about the difference between the quality of features and services that an app offers on different platforms and devices. Users often find such differences unfair and expect to be treated equally by app providers. In the following review, an Android user feels that they are not being treated equally as they receive the updates later than iOS users and have videos with less quality despite having a higher-quality camera than iOS users.

\faThumbsODown{} \textit{``I love the idea of [the app name], but I hate how biased it is towards Apple products. As an Android phone owner, my camera has more megapixels than an iPhone but my videos come out looking horribly pixelated. Android users also receive updates last and it is not fair. As customers, we deserve the same treatment that Apple users get. Make [the app] equal.''}

In another example, a user thinks that Samsung phone users face difficulties to receive the same features that iPhone users have had for a while because they have to buy the latest Samsung phone.

\faThumbsODown{} \textit{``[The app name] annoys me because of how unfair it is. [It] does not give any good updates to Samsung phones as it does to iPhone. Samsung users have to get the latest Samsung phone to get the new updates which most iPhone [users] that are years old have already got. I am one lower version down so I can't get the customized faces, which pretty much all versions of iPhone have. So don't treat us differently.''}

The feeling of inequality also happens when users of certain platforms and devices are deprived of some features and services that are available to the users of other platforms and devices. In the following example, an Android user feels that iOS users are given privileges as the Android version of the app does not provide the dark mode, while the iOS version does.

\faThumbsODown{} \textit{``I really love this app, I really do but, why does iOS have a dark mode, and Android doesn't? Dark mode would help so much as the light mode hurts my eyes. Please could we have dark mode?''}

\textbf{Fairness Concern 2: Feeling linguistic discrimination.} The main focus of this fairness concern is about the discrimination or bias that the users perceive from the apps towards specific languages. Our analysis of the reviews that include this type of fairness concern indicates the importance of language as one of the signs of identity and sense of belonging for the users. As shown in the following examples, the lack of support for their languages or bias toward this attribute makes them feel to be excluded by app providers. In the first example, one user requests that the app should support the Urdu language as a large number of Urdu speakers in different countries cannot use the app. A user in the second review example raises more or less the same concern as the app supports all languages except Arabic. 

\faThumbsODown{} \textit{``Please add Urdu language support. There are many in India, Pakistan and the gulf countries who are only able to read/write Urdu.''}

\faThumbsODown{} \textit{``Where is the Arabic language? All languages except Arabic.''}

\textbf{Fairness Concern 3: Lack of transparency and fairness in dealing with user-generated content.} This fairness concern mainly comes from users of apps that belong to the `social' and `video players \& editors' categories. Some reviews criticize apps and their employed algorithms as they unfairly and non-transparently treat user-generated content (e.g., videos, images). In the following review, the user complains that despite having created quality videos, their videos are less viewed and liked than non-quality videos produced by others.  

\faThumbsODown{} \textit{``Suddenly [the app] stopped views on my videos. In spite of having 38k followers I'm not getting 200 views ... Best videos get no views while 3rd class videos have views in millions.''}

In the following review, a creator in a `video playing and editing' app compares themselves with another counterpart and feels that the app treats their generated video differently.

\faThumbsODown{} \textit{``I've been an editor on [the app name] ... and it's really frustrating to have my views, comments, likes and subscribers get deleted. I work hard for my channel and getting this result is really disappointing. This is pretty unfair for me as my editor friends already crossed 10k views on at least one video. Please do something. I am only a small creator and not even a big one so each and every thing means a lot.''}

\textbf{Fairness Concern 4: Feeling gender and racial discrimination.} Another fairness concern that we identified in app reviews is that some apps seem to favour or be biased toward one gender. We also found that some apps exclude or ban some users based on their gender and race. In the reviews below, the users feel inequality as the apps are not gender-inclusive.

\faThumbsODown{} \textit{``I don't know about the concept used behind this app but my ques is this app made solely for girls? All the filters and modes [are] available only for girls!''}

\faThumbsODown{} \textit{``This app is very racist it doesn't let me choose my sexuality and won't accept it. I would also like to be able to have people from my own country but the app won't let me''.}

Below is an example of some reviews that complain some apps block some user accounts based on their race.

\faThumbsODown{} \textit{``This app targeted my race all because of how I look they banned me for no reason. I feel like my rights are taken from me and I think that is very wrong. So if you download this be very careful every one on this app will judge you it's not a fun app at all.''}

\textbf{Fairness Concern 5: Feeling biased censorship and promotion.} Our exploration into fairness reviews led to finding another fairness concern in which users perceive biased censorship and promoting of a certain group of people as a form of unfair behaviour that violates their free speech rights and marginalizes their views. We found that most such biases are driven by political views. Here are two examples of such reviews. In the first review, the user believes that the algorithms employed by the app are manipulated for political purposes.

In the second review, while the user admires the functionality of the app, they slightly accuse the app of having biased behaviour toward opposing political views.

\faThumbsODown{} \textit{``Unjust demonetization, manipulation of algorithms to push political correctness and censorship of free speech.''}

\faThumbsODown{} \textit{``The app works fine but is pretty intolerant and biased toward opposing political views.''}

\textbf{Fairness Concern 6: Unfair and non-transparent advertisement and subscription policies.} This fairness concern reflects the complaints of users about unfair, non-transparent, and sometimes misleading policies adopted by some apps for advertising and subscription. Some users state that while they are paying for a premium version and supposedly they should not receive any advertisements (e.g., it is mentioned in the advertisement policy), they are still receiving in-app advertisements.

\faThumbsODown{} \textit{``[I am] extremely disappointed. I pay for [the premium version] and I can only skip a couple of times then after that it does not let me stop any of my stations. It is the exact same way when you use [the free version] on a PC. I thought when you pay you get better service but guess not. I pay 4 dollars each month just to get rid of ads.''}

We also found some reviews where users feel that no freedom left for them from the app and they are pushed to pay and subscribe for the premium version. In the following example, the user states that the app shows an overwhelming number of advertisements to force them to subscribe to a premium version. 

\faThumbsODown{} \textit{``Too many ads! I don't know what changed, but I'm getting 2 back to back ads pushing me to pay for a subscription for every 2 songs I listen to. All you did is force me to subscribe to [the app name] for double the price out of spite.''}

We have also seen some reviews where the users imply that the cost of the premium service in some apps is not fair and reasonable for those people who cannot afford the subscription, consequently that they are excluded from enjoying the benefits of the service.

\faThumbsODown{} \textit{``So many ads! Yes, I can't afford your subscription but I am fed up with your ads.''}

\begin{tcolorbox}[arc=0mm,width=1.02\columnwidth,
                  top=0mm,left=0mm,  right=0mm, bottom=0mm,
                  boxrule=.75pt]
\textbf{Key Finding 2}: Our analysis reveals that users discuss six types of fairness concerns in AI-based app reviews, among which ‘\textit{receiving different quality of features and services in different platforms and devices}’ and ‘\textit{lack of transparency and fairness in dealing with user-generated content}’ are the most widely raised fairness concerns.
\end{tcolorbox}

\section{Root Causes of Fairness Concerns (RQ3)}\label{sec:rootcause}

\subsection{Approach}\label{sec:approachreactions}
As shown in Fig. \ref{fig:binarykmeans}, our model identified 91,974 fairness reviews (Section \ref{sec:binaryresults}), which were clustered into six types of fairness concerns (Section \ref{sec:sumresults}). These 91,974 app reviews received 4,911 responses from app owners. To understand the root causes reported by app owners for the identified fairness concerns, we randomly selected 2,248 out of the 4,911 app owners' responses to fairness reviews with a confidence level of 99\% and margin error of 2\%~\cite{israel1992determining, kadam2010sample}. In the first step, the first author conducted the open coding procedure~\cite{glaser2017discovery} on these 2,248 app owners' responses. Next, the third author reviewed all the codes and categories produced by the first author. Last, the first and third authors held several meetings to discuss and refine the codes and categories. Similar to Section \ref{sec:dataset}, the negotiated agreement approach \cite{campbell2013coding} was employed to resolve any disagreements and conflicts between the authors. This process led to the identification of six root causes of fairness concerns.

\subsection{Results} \label{sec:rootcauseresult}
This section describes the six root causes of fairness concerns, based on our analysis of the app owners' responses.

\textbf{1. Copyright Issues.} App owners admit that some of the fairness concerns are due to the copyright issues imposed on the apps, platforms, or devices. This can mainly be a reason behind Fairness Concern 1 (receiving different quality of features and services in different platforms and devices) identified in RQ2 (Section \ref{sec:sumresults}). For example in the following conversation, the copyright issue is mentioned by an app owner as the reason why Android and iOS users do not receive the same emojis.

\faThumbsODown{} \textit{``This is good but I wanted iPhone emojis, not the Android ones or whatever it is. It would be better if it was changed to the iPhone one.}

\faReply{} \textit{``We understand your concern but there’s some copyright issue with iOS. If you are using an Android device, then you will send Android version of emojis so far. If one is using an iOS device, then he/she will automatically receive iOS version of emojis. It really depends on the device one is using. Hope it helps!''}

\textbf{2. Development Complexity.} 
As discussed in Section \ref{sec:sumresults}, one of the complaints raised by users in Fairness Concern 1 is that Android and iOS users do not receive new features and/or updates simultaneously. Our analysis of app owners' responses found that users experience this seemingly because the complexity of developing certain features differs among Android and iOS. Hence, the development and deployment of a feature may need more time on one platform than on others. 

\faThumbsODown{} \textit{``Why are the filters only for iPhone?''}

\faReply{} \textit{``[This is] because Android filters are more complicated to develop, so we are still working hard on it. Sorry about this!''}

\textbf{3. \textcolor{black}{Faults}.} 
\textcolor{black}{Some app owners admit that faults in AI-based apps could contribute to the identified fairness concerns (particularly Fairness Concerns 4 and 5). For example, in the following response, the owner accepted that an error may have led to incorrectly banning the user.}

\faThumbsODown{} \textit{``I got banned from the app for no reason at all. I really think it is unfair.''}

\faReply{} \textit{``You may have been banned by mistake. Please try logging in again. If you're still experiencing issues, please contact us at...''}

In response to a user who criticizes a chatbot app for producing racist messages to a specific ethnic group, the app owner admits that the used AI models may make mistakes in identifying the context of a conversation, leading to saying inappropriate messages. 

\faReply{} \textit{``Sometimes, due to the nature of AI models, [app name] can generate responses that are untrue when it thinks it will be relevant to the conversation. [App name] is not a sentient being and can play along with anything you come up with. To avoid chats like that, downvote the messages from your [app name] or mark them offensive!''.}

\textbf{4. External Factors.} 
In justifying the fairness concerns raised by users (particularly Fairness Concerns 2, 5, and 6), app owners sometimes refer to factors that are out of their control. These external factors are diverse and range from rules, policies, and cultural norms that apps should follow (in a country) to restrictions imposed by platforms and infrastructures. In the following response, the app owner states that the user cannot see Gujarati-language characters, not because their app does not support Gujarati language but instead due to the lack of support from the user's device.

\faReply{} \textit{``Sorry you are experiencing this issue. This is unrelated to [App Name] - it is related to the type of device you are using. If your device supports the Gujarati language font, then you will be able to see the characters on your keyboard.''}

In another example, an app owner replies to the user that they cannot fix a reportedly unfair and non-transparent subscription payment as payment processes are managed by a third party.

\faThumbsODown{} \textit{``Misleading advertising for services in-app subscription. [It] says \$6.99 a month/cancel anytime then they charge nearly \$75 for a whole year.''}

\faReply{} \textit{``We do not manage or handle subscription payments or payment processing \& are unable to help you with a refund. Please reach out to Google Support for help!''}

\textbf{5. Development Cost.} In Fairness Concern 6, we found that some users feel that the number of in-app advertisements and/or the price of subscriptions is unfair. Our analysis found that app owners mainly point out the `cost' associated with developing and maintaining the apps as a reason to respond to such a fairness concern. The following are the replies of two app owners to the users who complain the subscription is too costly for kids and third-world countries, respectively.

\faThumbsODown{} \textit{``I don't get why you have to pay for this. Some of us are kids and don't have a lot of money. I know you can have it for free but it's only for 7 days not all of us are rich/adults. I was really hoping I could at least search a song but no, that's also premium I don't really know why. Can you at least make the search not premium? I don't want to play the ones they give me, I want to play what I want but I can't cause I have to pay to search for songs.''}

\faReply{} \textit{``We have a small team that works very hard to continuously provide updates to the app. Unfortunately, it is very expensive to develop and maintain an app, and providing you with hundreds of popular songs does not come cheap. I hope you decide to give us a chance in the future!''}

\vspace*{1em}
\faThumbsODown{} \textit{``The subscription is too costly for third-world countries.''}

\faReply{} \textit{``We are trying to make subscriptions as affordable for all users as possible, but we still need them to keep working on our app and continue developing it.''} 

\textbf{6. User Usage and Awareness.} Our analysis reveals that the fairness concerns in some cases are mistakenly raised by users due to they either poorly follow app usage instructions or have limited awareness of app features and capabilities. In the following dialogue, a user feels it is not possible to unsubscribe from an app. This can be considered as unfair and non-transparent subscription policy (Fairness Concern 6) from the user perspective. However, the app owners emphasizes that users are able to do so anytime and refer the subscription instruction.

\faThumbsODown{} \textit{``Once you subscribe you can't unsubscribe so now I'm paying 3\$ a month in perpetuity.''}

\faReply{} \textit{``You definitely can unsubscribe at any moment. Please follow this instruction to cancel or change your subscription.''}

In the conversation below, while a user feels linguistic discrimination (Fairness Concern 2) as they think the app does not provide translation to/from the Hausa language, the app owner informs the user that their app provides some Hausa translation features.

\faThumbsODown{} \textit{``Why is it that you are not translating Hausa the most popular language in Africa? or is it because you don't like Africans?''}

\faReply{} \textit{``[App name] supports some Hausa translation features. You can check the availability and supporting features for any language here ...''}

\begin{tcolorbox}[arc=0mm,width=1.02\columnwidth,
                  top=0mm,left=0mm,  right=0mm, bottom=0mm,
                  boxrule=.75pt]
\textbf{Key Finding 3}: App owners report six root causes, namely `copyright issues', `development complexity', `buggy code', `external factors', `development cost', and `user usage and awareness', to justify the fairness concerns raised by users.
\end{tcolorbox}

\input{Tables/FairnessConcernsInCategories}

\section{Discussion} \label{sec:discussion}
In this section, we reflect on our findings and some implications (indicated with the \faLeanpub \hspace{0.5mm} icon) for research and practice. 

\textbf{\textit{Relationship between fairness concerns and app categories.}} \textcolor{black}{As discussed in Section~\ref{sec:distribution}, the proportion of fairness reviews varies among app categories. While fairness reviews are more common in the `communication' and `social' categories, the users in the `personalization' and `auto \& vehicles' categories rarely discuss fairness concerns in app reviews (see Table \ref{tbl:fairnessincategories}). We also found that the frequency of the six fairness concerns identified in Section~\ref{sec:sumresults} differs in app categories. As shown in Table \ref{tbl:clustergenres}, \textbf{Fairness Concerns 1}, \textbf{3}, \textbf{4}, and \textbf{5} are frequently discussed by users in the `social' category. Among the identified concerns, \textbf{Fairness Concern 1} is pervasive in the `communication' category, followed by \textbf{Fairness Concern 4}. On the other hand, \textbf{Fairness Concern 3}, which refers to the lack of transparency and fairness in dealing with user-generated content, is prevalent in apps that come from the `social' and `video players \& editors' categories, while it is rare in the `lifestyle', `art \& design', and `auto \& vehicles' categories. \textbf{Fairness Concern 6}, referring to unfair and non-transparent advertisement and subscription policies, are mainly discussed in the `video players \& editors' and `music \& audio' categories.}
\faLeanpub \hspace{0.5mm} These observations collectively indicate that there might be a correlation between app categories (the nature of apps) and the types of fairness concerns. We suggest future research efforts to investigate this relationship and delve deeper into the fairness concerns specific to each category. We posit that the findings of such focused research can help app development organizations allocate specialized and customized resources and mechanisms to guide and train developers to address the most important and frequent fairness concerns in each app category.    

\textbf{\textit{Fairness is a socio-technical concern.}} AI-based apps may include stakeholders such as users, developers, app owners, policymakers, and ethics experts. Our study exclusively explored fairness concerns in AI-based mobile apps from the users' perspective. Note that not all the six fairness concerns considered in our study have to do with AI components of the studied AI-based apps. For example, while users in \textbf{Fairness Concern 6} mainly accuse apps due to unfair policies and decisions taken by the app providers, we have seen that users sometimes explicitly talk about the seemingly unfair AI algorithms employed by apps in \textbf{Fairness Concerns 3} and \textbf{5}. The analysis of app owners' responses to the fairness reviews in RQ3 shows that app owners also report several non-technical causes to justify the identified fairness concerns. For instance, while many Android users feel that they are not treated equally as iOS users in the apps that are available in both Android and iOS versions (see \textbf{Fairness Concern 1}), we found that Android app owners point out `policy issues' and `development complexity' as the root causes of this fairness concern, which can prevent them from providing the exactly same services and features to Android users. \faLeanpub \hspace{0.5mm} In line with the findings of \cite{ekstrand2022fairness,lu2022towards,selbst2019fairness}, we re-emphasize that fairness is a socio-technical concern that may stem from technical components such as algorithms or any other aspects of AI-based systems such as policies adopted by a company.



\textbf{\textit{Root causes of fairness concerns in AI-based apps.}} In RQ3, we identified six root causes that app owners report to justify the fairness concerns raised by users. We observed that app owners not only admit technical root causes (e.g., `buggy code') behind the raised fairness concerns but also refer to different limitations and non-technical challenges (e.g., `policy issues' and `external factors') in this regard. Furthermore, our observation in RQ3 shows that app owners either do not reply to the majority of the fairness reviews or frequently use a predefined template (e.g., to apologise to users) to respond to users. This is aligned with the findings by Chen et al. \cite{chen2021should}, as they found that app owners rarely attempted to justify user interface-related issues reported by users in mobile app reviews. Hence, analyzing app owners' replies to the fairness reviews may fail to identify all types of root causes behind fairness concerns. In alignment with the findings in \cite{holstein2019improving,green2018myth}, we also observed that the formation of user subgroups within AI-based systems is not limited to well-established attributes such as gender or language. Intriguingly, there are certain subgroups that are highly contingent on context and application, making them particularly challenging for developers to identify. For example, in \textbf{Fairness Concern 3}, users in the `social' and `video players \& editors' categories often compared themselves to different types of creators, such as high-profile creators versus ordinary users. However, we rarely found any concrete responses from app owners to such fairness concerns. \faLeanpub \hspace{0.5mm} Given that AI-based apps include several types of stakeholders and the potential limitations of analyzing app owners' responses, we argue that more research should be conducted with a wide range of stakeholders in AI-based apps to identify a comprehensive list of the root causes of fairness concerns in each app category.

\textbf{\textit{Benefits of findings for mobile app development tasks.}} In a systematic review, Dąbrowski et al. \cite{dkabrowski2022analysing} found that app review analysis can support various software development activities (e.g., requirement engineering and design). We argue that the findings obtained in RQ2 (fairness concerns from the users' perceptive) and RQ3 (root causes of fairness concerns from the app owners' viewpoint) could benefit mobile app development in different ways. (1) The identified fairness concerns and their potential root causes can help app owners and developers understand and elicit users' fairness-related requirements and expectations. For example, the identified fairness concerns can guide app developers on questions that should be asked of users to gather fairness-related requirements and expectations. (2) The knowledge of potential fairness concerns can help app developers be aware of and avoid common fairness concerns during app development. (3) The identified fairness concerns can act as a checklist for app developers to enable them to specify test scenarios that evaluate specific fairness concerns. (4) Our findings can inform researchers and tool builders to develop tools to embed fairness in AI-based apps, and detect and fix fairness concerns. \textcolor{black}{(5) The six identified fairness concerns could serve as the basis for constructing multi-label datasets of fairness concerns and developing multi-classification techniques that automatically detect these fairness concerns in text-based software artifacts.}


\textbf{\textit{Applications and limitations of the employed methodology.}} \textcolor{black}{Our methodology in RQ1 consists of developing a set of learning models that distinguish fairness reviews from non-fairness reviews. The clustering technique in RQ2 uses the fairness reviews detected by the best model as input to summarize fairness concerns.} We also developed a fairness keyword set using an iterative and multi-faceted approach to find potential fairness reviews. This keyword set helped significantly reduce the efforts of creating the ground-truth dataset as approximately 50\% of potential fairness reviews chosen for the manual labeling were finally labeled as fairness reviews by human experts. This benefit becomes more apparent in light of our observation that fairness reviews roughly constitute $\approx$1\% of all reviews in an app. \textcolor{black}{This iterative and multi-faceted approach to creating the ground-truth dataset gave us a nuanced understanding of the complex nature of fairness within the app ecosystem \cite{avelar22me}. As a result, we built a highly accurate dataset that served as the foundation for training models, achieving an impressive precision of 94\% in identifying fairness reviews.} \faLeanpub \hspace{0.5mm} Other researchers may apply this iterative and multi-faceted approach to reduce false positives, particularly when working with infrequently discussed topics in app reviews. In Section \ref{sec:fairnessclassifier}, we experimented with different models, from classical ML models to LLMs, to select the best-performing model. The performance of the models on the metrics was not consistent. Some achieved a good recall but had a very low precision, or vice versa. Hence, we had difficulties in choosing the best-performing model. We selected Logistic Regression with TF-IDF, USE, and Word2Vec as the best-performing model because it had the highest precision and AUC. This decision significantly helped us to reduce the noises in the clusters in Section \ref{sec:cluster}. \faLeanpub \hspace{0.5mm} \textcolor{black}{We argue that the popularity of pre-trained and large language models should not result in less attention being paid to traditional ML models, which may outperform DL models in some datasets.}

\section{Threats to Validity} \label{sec:validity}

\textbf{Internal Validity}. A key concern in our internal validity is the inherent ambiguity attached to the concept of `fairness', which can be nuanced, with different interpretations. App users might express fairness concerns using various terms and phrases, not all of which are easy to capture by a keyword-based approach like ours. To mitigate the potential impact of this threat, we used an iterative and multi-faceted approach to identify fairness-related keywords. We started with a literature-driven set of keywords and expanded this using KeyBERT. The keyword set was reviewed and refined manually to ensure relevance and accuracy. \textcolor{black}{Moreover, our manual labeling process in RQ1 and qualitative analysis in RQ2 and RQ3 included multiple coders to minimize individual bias and capture a broader interpretation of fairness.}

\textbf{Conclusion Validity}. The conclusions regarding the fairness concerns of users and the root causes of the fairness concerns were based on the results of our clustering approach. \textcolor{black}{As discussed in Section \ref{sec:cluster}, we excluded Clusters 0 and 1 as they were not compact clusters. These two clusters together contained around 18,090 reviews. Such exclusion could have led to missing some important types of fairness concerns. Furthermore, we only experimented with $k$=2 to 10 to detect the optimum number of clusters. We acknowledge that investigating a larger $k$ may lead to identifying more or different fairness concerns, particularly the long-tail of fairness concerns. We also concede
that different instrumentation of our clustering approach might have influenced our conclusions in Section~\ref{sec:sumresults} and Section~\ref{sec:rootcauseresult}.} To mitigate the potential impact of this threat, we strived for transparency and thoroughness in our reporting. We explained our choices in the clustering process, including why a particular technique was selected and our parameter-tuning process. We also acknowledge the influence of these choices on our findings. While our choices were reasonable and based on best practices, further experimentation with more techniques is required in our future work. \textcolor{black}{Finally, Table \ref{tbl:AIApps} illustrates that 7 categories had only one representative app in our analysis. This may have impacted our conclusions on fairness concerns in app categories (see Table \ref{tbl:clustergenres}). Future research could build upon our work by expanding our dataset to include a larger and equal number of apps in each category, enhancing the findings' reliability.}  

\textbf{External Validity}. Our evaluation is based on 108 apps from 23 categories and an overall analysis of $\approx$9.5 million reviews. The results we received across these apps with a thorough manual analysis of $>$2,600 reviews for building our model, analysis of $>$92,000 reviews using the model and clustering, and qualitative analysis of 2,248 lend a degree of confidence to the generalizability of our study. Further experimentation is nonetheless required to mitigate external validity threats.


\section{Conclusion and Future Work} \label{sec:conclusion}
In this study, we analyzed the reviews of AI-based apps to identify the types of fairness concerns users experience when using AI-based apps. We also investigated root causes reported by app owners to justify the fairness concerns. Our analysis was based on $\approx$9.5M reviews collected from 108 AI-based apps. Given that app reviews cover different topics, we first developed a set of ML and DL models to identify reviews that include discussions around fairness (i.e., fairness reviews). We used a manually constructed dataset of fairness and non-fairness reviews to train and evaluate these ML and DL models. We found that Logistic Regression with TF-IDF, USE, and Word2Vec as the feature extraction technique reaches the highest precision of 0.94. Furthermore, our experiments with K-mean clustering on $\approx$92K fairness reviews resulted in six types of fairness concerns in AI-based apps. Our qualitative analysis of 2,248 app owners' responses identified six root causes (e.g., `policy issues', `development cost') of the fairness concerns.

We plan to conduct further studies with other stakeholders of AI-based apps such as app owners, developers, and policymakers, and obtain more data from other sources (e.g., GitHub) to generalize our understanding of fairness concerns in AI-based apps. \textcolor{black}{We also plan to experiment with other clustering techniques and evaluate the extent to which the employed learning models adhere to ethical principles, with a particular focus on fairness.}

\begin{acks}
This work has been partially supported by the National Natural Science Foundation of China (NSFC) with Grant No. 62172311.
\end{acks}

\bibliographystyle{ACM-Reference-Format}
\bibliography{basebib}










\end{document}

%% file: Tables/AIapps.tex
\begin{table}[]
\caption{\textcolor{black}{The number of the identified AI-based apps in each category.}}
\label{tbl:AIApps}
\small
\setlength{\tabcolsep}{4pt}
\renewcommand{\arraystretch}{1.1}
\begin{tabular}{|l|l|l|l|l|l|}
\hline
\textbf{\#} & \textbf{Categories}      & \textbf{Number of AI Apps} & \textbf{\#} & \textbf{Categories} & \textbf{Number of AI Apps} \\ \hline
1           & Photography              & 15                         & 13          & Health \& Fitness   & 3                          \\ \hline
2           & Productivity             & 13                         & 14          & Food \& Drink       & 3                          \\ \hline
3           & Communication            & 10                         & 15          & Lifestyle           & 3                          \\ \hline
4           & Shopping                 & 10                         & 16          & Role Playing        & 2                          \\ \hline
5           & Tools                    & 9                          & 17          & Arcade              & 1                          \\ \hline
6           & Social                   & 7                          & 18          & Education           & 1                          \\ \hline
7           & Video Players \& Editors & 6                          & 19          & Strategy            & 1                          \\ \hline
8           & Travel \& Local          & 5                          & 20          & Art \& Design       & 1                          \\ \hline
9           & Entertainment            & 4                          & 21          & Auto \& Vehicles    & 1                          \\ \hline
10          & Finance                  & 4                          & 22          & Puzzle              & 1                          \\ \hline
11          & Business                 & 4                          & 23          & Personalization     & 1                          \\ \hline
12          & Music \& Audio           & 3                          &             &                     &                            \\ \hline
\end{tabular}
\end{table}

%% file: Tables/metrics.tex
\begin{table}[]
\caption{\textcolor{black}{The metrics used to evacuate the models. TP: True Positive, FP:False Positive, TN: True Negative, FN: False Negative, TPR: True Positive Rate, FPR: False Positive Rate}}
\label{tbl:Metrics}
\renewcommand{\arraystretch}{1.4}
\begin{tabular}{|l|l|m{8cm}|}
\hline
\textbf{Metrics} & \textbf{Calculations}                           & \textbf{Descriptions}                                                                                                                                                                                                                                                                                          \\ \hline
Precision        & $\frac{TP}{TP+FP}$                              & \textcolor{black}{It is the proportion of correctly predicted fairness reviews out of all reviews predicated as fairness reviews.}                                                                                                                                                                                          \\ \hline
Recall           & $\frac{TP}{TP+FN}$                              & \textcolor{black}{It is the proportion of correctly predicted fairness reviews out of the total number of fairness reviews.}                                                                                                                                                                                                \\ \hline
F1-score         & $\frac{2*Precision*Recall}{Precision + Recall}$ & \textcolor{black}{It combines precision and recall into one single score.}                                                                                                                                                                                                                                                  \\ \hline
Accuracy         & $\frac{TP+TN}{TP+FP+TN+FN}$                     & \textcolor{black}{It is the ratio of precisely predicted fairness reviews and non-fairness reviews to all reviews.}                                                                                                                                                                                                \\ \hline
AUC              & $\int_{0}^{1} TPR(t_{i}), dFPR(t_{i})$          & \textcolor{black}{It measures the likelihood of a model ranking a randomly selected fairness review (i.e., True Positive Rate) higher than a randomly selected non-fairness review (i.e., False Positive Rate). The AUC provides an aggregate measure of performance across all possible model thresholds (i.e., $t_{i}$).} \\ \hline
\end{tabular}
\end{table}

%% file: Tables/binaryclassifications.tex
\begin{table}[]
\caption{Results of the models (in \%). The best result in each metric is greyed and bold. The model chosen as the best-performing model is bold and yellowed. P: Precision, R: Recall, F1: F1-score, ACC: Accuracy, AUC: Area under the ROC Curve, XGB: XGBoost, W2V: Word2Vec.}
\label{tbl:MLResults}
\small
\setlength{\tabcolsep}{4pt}
\renewcommand{\arraystretch}{1.3}
\begin{tabular}{|l|l|c|c|c|c|c|}
\hline
& \textbf{Models}   & \textbf{P} & \textbf{R} & \textbf{F1} & \textbf{ACC} & \textbf{AUC} \\ \hline
\multirow{21}{*}{Traditional ML Models} & LR+TFIDF          & \cellcolor[HTML]{EFEFEF}\textbf{88.23}      & 48.78      & 62.50       & 74.78        & 88.18        \\ \cline{2-7}
& LR+USE            & 83.80      & 63.71      & 72.29       & 78.89        & 89.04        \\ \cline{2-7}
& LR+W2V            & 80.38      & 48.71      & 60.23       & 72.40        & 83.86        \\ \cline{2-7}
& LR+TFIDF+USE      & 85.07      & 67.10      & 74.93       & 80.69        & 90.35        \\ \cline{2-7}
& LR+TFIDF+W2V      & 84.64      & 58.59      & 68.99       & 77.39        & 87.85        \\ \cline{2-7}
& LR+USE+W2V        & 83.47      & 66.93      & 74.18       & 79.96        & 89.45        \\ \cline{2-7}
& \cellcolor[HTML]{FFE135}\textbf{LR+TFIDF+USE+W2V}  & 84.99      & 70.02      & 76.61       & 81.61        & \cellcolor[HTML]{EFEFEF}\textbf{90.37}        \\ \cline{2-7}
& SVM+TFIDF         & 80.75      & 70.52      & 70.08       & 79.92        & 87.60        \\ \cline{2-7}
& SVM+USE           & 81.64      & 69.66      & 75.15       & 80.12        & 88.61        \\ \cline{2-7}
& SVM+W2V           & 78.75      & 57.92      & 66.47       & 74.86        & 83.83        \\ \cline{2-7}
& SVM+TFIDF+USE     & 83.58      & 73.95      & 78.31       & \cellcolor[HTML]{EFEFEF}\textbf{82.53}        & 90.28        \\ \cline{2-7}
& SVM+TFIDF+W2V     & 81.80      & 69.73      & 75.14       & 80.15        & 87.99        \\ \cline{2-7}
& SVM+USE+W2V       & 81.12      & 70.53      & 75.39       & 80.19        & 88.98        \\ \cline{2-7}
& SVM+TFIDF+USE+W2V & 82.52      & 74.30      & 78.01       & 82.11        & 90.09        \\ \cline{2-7}
& XGB+TFIDF         & 77.90      & 74.05     & 75.74       & 79.54        & 87.61        \\ \cline{2-7}
& XGB+USE           & 80.65      & 74.67      & 77.35       & 81.11        & 89.11        \\ \cline{2-7}
& XGB+W2V           & 76.89      & 71.24      & 73.73       & 78.04        & 86.26        \\ \cline{2-7}
& XGB+TFIDF+USE     & 80.68      & 75.17      & 77.71       & 81.35        & 89.93        \\ \cline{2-7}
& XGB+TFIDF+W2V     & 78.66      & 74.61      & 76.38       & 80.08        & 88.38        \\ \cline{2-7}
& XGB+USE+W2V       & 80.70      & 76.28      & 78.35       & 81.77        & 89.59        \\ \cline{2-7}
& XGB+TFIDF+USE+W2V & 81.01      & 75.70      & 78.17       & 81.73        & 89.93        \\ \hline
\multirow{2}{*}{Pre-trained Language Models}& BERT              & 82.38      & 77.08      & 79.63       & 80.29        & 80.29        \\ \cline{2-7}
& RoBERTa           & 84.99      & 78.31      & \cellcolor[HTML]{EFEFEF}\textbf{81.50}       & 82.24        & 82.25        \\ \hline
Large Language Models& GPT-3             & 63.18      & \cellcolor[HTML]{EFEFEF}\textbf{81.46}      & 71.16       & 71.28        & 72.46        \\ \hline
\end{tabular}
\vspace*{-1em}
\end{table}

%% file: Tables/unseentable.tex
\begin{table}[]
\caption{\textcolor{black}{Results of best-performing model on a sample of all reviews dataset}}\label{unseentab}
\begin{tabular}{|lccccc|}
\hline
\multicolumn{1}{|l|}{}       & \multicolumn{1}{c|}{P}     & \multicolumn{1}{c|}{R}     & \multicolumn{1}{c|}{F1}     & \multicolumn{1}{c|}{ACC}   & AUC   \\ \hline
\multicolumn{1}{|l|}{LR+TDIDF+USE+W2V} & \multicolumn{1}{c|}{88.75} & \multicolumn{1}{c|}{95.17} & \multicolumn{1}{c|}{91.85} & \multicolumn{1}{c|}{92.12} & 92.32 \\ \hline
\end{tabular}
\end{table}

%% file: Tables/fairnessincategories.tex
\begin{table}[]
\caption{Number/percentage of fairness reviews per category}
\label{tbl:fairnessincategories}
\small
\setlength{\tabcolsep}{4pt}
\renewcommand{\arraystretch}{1.3}
\begin{tabular}{|l|l|l|l|}
\hline
\textbf{Categories}      & \textbf{\begin{tabular}[c]{@{}l@{}}All reviews\end{tabular}} & \textbf{\begin{tabular}[c]{@{}l@{}}Fairness reviews\end{tabular}} & \textbf{\begin{tabular}[c]{@{}l@{}}\% of fairness reviews\end{tabular}} \\ \hline
Communication            & 1,610,525                                                                   & 28,304                                                                          & 1.76\%                                                                             \\ \hline
Social                   & 1,435,621                                                                   & 23,602                                                                          & 1.64\%                                                                             \\ \hline
Entertainment            & 214,384                                                                    & 3,118                                                                           & 1.45\%                                                                             \\ \hline
Arcade                   & 33,066                                                                     & 364                                                                            & 1.10\%                                                                             \\ \hline
Role Playing             & 13,5610                                                                    & 1,373                                                                           & 1.01\%                                                                             \\ \hline
Video Players \& Editors & 1,008,512                                                                   & 9,583                                                                           & 0.95\%                                                                             \\ \hline
Music \& Audio           & 554,215                                                                    & 4,766                                                                           & 0.86\%                                                                             \\ \hline
Travel \& Local          & 177,341                                                                    & 1,384                                                                           & 0.78\%                                                                             \\ \hline
Tools                    & 937,639                                                                    & 6,823                                                                           & 0.73\%                                                                             \\ \hline
Health \& Fitness        & 152,649                                                                    & 898                                                                            & 0.59\%                                                                             \\ \hline
Food \& Drink            & 192,842                                                                    & 10,087                                                                          & 0.56\%                                                                             \\ \hline
Art \& Design            & 47,715                                                                     & 249                                                                            & 0.52\%                                                                             \\ \hline
Education                & 47,464                                                                     & 214                                                                            & 0.45\%                                                                             \\ \hline
Productivity             & 906,605                                                                    & 3,645                                                                           & 0.40\%                                                                             \\ \hline
Finance                  & 84,209                                                                     & 332                                                                            & 0.39\%                                                                             \\ \hline
Strategy                 & 51,970                                                                     & 203                                                                            & 0.39\%                                                                             \\ \hline
Photography              & 1,095,201                                                                   & 3,888                                                                           & 0.35\%                                                                             \\ \hline
Lifestyle                & 82,604                                                                     & 282                                                                            & 0.34\%                                                                             \\ \hline
Shopping                 & 505,533                                                                    & 1,547                                                                           & 0.31\%                                                                             \\ \hline
Puzzle                   & 11,310                                                                     & 26                                                                             & 0.23\%                                                                             \\ \hline
Business                 & 160,373                                                                    & 258                                                                            & 0.16\%                                                                             \\ \hline
Personalization          & 19,390                                                                     & 19                                                                             & 0.09\%                                                                             \\ \hline
Auto \& Vehicles         & 10,727                                                                     & 9                                                                              & 0.08\%                                                                             \\ \hline
\end{tabular}
\vspace*{-1em}
\end{table}

%% file: Tables/FairnessConcernsInCategories.tex
\begin{table*}[]
\caption{\textcolor{black}{Frequency of fairness concerns (clusters) in each category. FC: Fairness Concern }}
\label{tbl:clustergenres}
\small
\setlength{\tabcolsep}{4pt}
\renewcommand{\arraystretch}{1.1}

\begin{tabular}{|l|c|c|c|c|c|c|}
\hline
\textbf{Categories}      & \textbf{FC1}                   & \textbf{FC2}                  & \textbf{FC3}                  & \textbf{FC4}                  & \textbf{FC5}                  & \textbf{FC6}                  \\ \hline
Photography              & \cellcolor[HTML]{BCD698}492    & \cellcolor[HTML]{F2F6E8}63    & \cellcolor[HTML]{ECF3DF}84    & \cellcolor[HTML]{BAD597}571   & \cellcolor[HTML]{EAF1DB}95    & \cellcolor[HTML]{BED699}442   \\ \hline
Productivity             & \cellcolor[HTML]{BFD699}417    & \cellcolor[HTML]{C3D79B}286   & \cellcolor[HTML]{FAFCF6}28    & \cellcolor[HTML]{C2D79A}334   & \cellcolor[HTML]{EAF1DC}93    & \cellcolor[HTML]{C3D79B}304   \\ \hline
Communication            & \cellcolor[HTML]{00B050}11,942 & \cellcolor[HTML]{B5D495}716   & \cellcolor[HTML]{B9D597}576   & \cellcolor[HTML]{86CB83}2,097 & \cellcolor[HTML]{BAD598}550   & \cellcolor[HTML]{C1D79A}362   \\ \hline
Shopping                 & \cellcolor[HTML]{FAFCF7}27     & \cellcolor[HTML]{DEE8C6}146   & \cellcolor[HTML]{F2F7E9}60    & \cellcolor[HTML]{C3D79B}308   & \cellcolor[HTML]{E6EED4}112   & \cellcolor[HTML]{C3D79B}302   \\ \hline
Tools                    & \cellcolor[HTML]{A3D18E}1,247  & \cellcolor[HTML]{A3D18F}1,222 & \cellcolor[HTML]{E7EFD6}107   & \cellcolor[HTML]{BCD698}500   & \cellcolor[HTML]{B0D394}862   & \cellcolor[HTML]{E4EDD0}121   \\ \hline
Social                   & \cellcolor[HTML]{68C578}2,969  & \cellcolor[HTML]{90CD87}1,794 & \cellcolor[HTML]{0BB254}5,706 & \cellcolor[HTML]{47BF6C}3,922 & \cellcolor[HTML]{48BF6C}3,904 & \cellcolor[HTML]{C0D79A}382   \\ \hline
Video Players \& Editors & \cellcolor[HTML]{D5E3B7}182    & \cellcolor[HTML]{C2D79B}313   & \cellcolor[HTML]{50C06F}3,656 & \cellcolor[HTML]{BAD597}572   & \cellcolor[HTML]{8DCC86}1,880 & \cellcolor[HTML]{98CF8A}1,555 \\ \hline
Travel \& Local          & \cellcolor[HTML]{F8FBF3}35     & \cellcolor[HTML]{EDF3E0}82    & \cellcolor[HTML]{F2F7E9}60    & \cellcolor[HTML]{CADBA4}229   & \cellcolor[HTML]{D1E0B1}197   & \cellcolor[HTML]{C7D9A0}240   \\ \hline
Entertainment            & \cellcolor[HTML]{C4D79B}250    & \cellcolor[HTML]{EBF2DD}89    & \cellcolor[HTML]{FCFDF9}20    & \cellcolor[HTML]{C7D9A0}239   & \cellcolor[HTML]{E5EED3}115   & \cellcolor[HTML]{DBE7C1}158   \\ \hline
Finance                  & \cellcolor[HTML]{FFFFFE}8      & \cellcolor[HTML]{FFFFFF}1     & \cellcolor[HTML]{FFFFFF}2     & \cellcolor[HTML]{FAFCF6}29    & \cellcolor[HTML]{FFFFFE}8     & \cellcolor[HTML]{DFEAC9}139   \\ \hline
Business                 & \cellcolor[HTML]{FFFFFF}5      & \cellcolor[HTML]{FEFFFD}11    & \cellcolor[HTML]{FFFFFF}1     & \cellcolor[HTML]{F6F9EF}45    & \cellcolor[HTML]{FFFFFE}9     & \cellcolor[HTML]{F5F8EE}49    \\ \hline
Music \& Audio           & \cellcolor[HTML]{CEDEAC}210    & \cellcolor[HTML]{ECF2DE}86    & \cellcolor[HTML]{F9FBF4}33    & \cellcolor[HTML]{C1D79A}364   & \cellcolor[HTML]{B2D494}788   & \cellcolor[HTML]{8DCC86}1,889 \\ \hline
Health \& Fitness        & \cellcolor[HTML]{FBFDF9}22     & \cellcolor[HTML]{FEFEFD}12    & \cellcolor[HTML]{FFFFFE}9     & \cellcolor[HTML]{D9E5BE}165   & \cellcolor[HTML]{ECF2DE}88    & \cellcolor[HTML]{C3D79B}283   \\ \hline
Food \& Drink            & \cellcolor[HTML]{CCDDA9}217    & \cellcolor[HTML]{FFFFFF}6     & \cellcolor[HTML]{FFFFFF}1     & \cellcolor[HTML]{DDE8C4}150   & \cellcolor[HTML]{F4F8ED}51    & \cellcolor[HTML]{E3ECCF}123   \\ \hline
Lifestyle                & \cellcolor[HTML]{FCFDF9}21     & \cellcolor[HTML]{FEFFFD}10    & \cellcolor[HTML]{FFFFFF}1     & \cellcolor[HTML]{F8FAF2}37    & \cellcolor[HTML]{FFFFFF}2     & \cellcolor[HTML]{F5F9EE}47    \\ \hline
Role Playing             & \cellcolor[HTML]{FFFFFF}4      & \cellcolor[HTML]{FFFFFF}6     & \cellcolor[HTML]{FDFEFB}17    & \cellcolor[HTML]{FCFDF9}20    & \cellcolor[HTML]{FDFEFB}15    & \cellcolor[HTML]{B8D597}607   \\ \hline
Arcade                   & \cellcolor[HTML]{FFFFFF}0      & \cellcolor[HTML]{FFFFFF}2     & \cellcolor[HTML]{FFFFFF}2     & \cellcolor[HTML]{F9FBF4}32    & \cellcolor[HTML]{FFFFFF}2     & \cellcolor[HTML]{D7E4BA}175   \\ \hline
Education                & \cellcolor[HTML]{FCFDFA}18     & \cellcolor[HTML]{FFFFFF}1     & \cellcolor[HTML]{FFFFFF}0     & \cellcolor[HTML]{FAFCF7}26    & \cellcolor[HTML]{FFFFFF}1     & \cellcolor[HTML]{E8F0D8}101   \\ \hline
Strategy                 & \cellcolor[HTML]{FFFFFF}3      & \cellcolor[HTML]{FFFFFF}3     & \cellcolor[HTML]{FFFFFF}3     & \cellcolor[HTML]{FEFFFD}11    & \cellcolor[HTML]{FFFFFF}2     & \cellcolor[HTML]{E7EFD6}107   \\ \hline
Art \& Design            & \cellcolor[HTML]{FDFEFB}15     & \cellcolor[HTML]{FFFFFF}1     & \cellcolor[HTML]{FFFFFF}1     & \cellcolor[HTML]{F2F6E9}61    & \cellcolor[HTML]{FFFFFF}0     & \cellcolor[HTML]{FAFCF6}28    \\ \hline
Auto \& Vehicles         & \cellcolor[HTML]{FFFFFF}0      & \cellcolor[HTML]{FFFFFF}0     & \cellcolor[HTML]{FFFFFF}1     & \cellcolor[HTML]{FFFFFF}1     & \cellcolor[HTML]{FFFFFF}0     & \cellcolor[HTML]{FFFFFF}1     \\ \hline
Puzzle                   & \cellcolor[HTML]{FFFFFF}1      & \cellcolor[HTML]{FFFFFF}0     & \cellcolor[HTML]{FFFFFF}0     & \cellcolor[HTML]{FFFFFF}0     & \cellcolor[HTML]{FFFFFF}0     & \cellcolor[HTML]{FFFFFF}2     \\ \hline
Personalization          & \cellcolor[HTML]{FFFFFF}5      & \cellcolor[HTML]{FFFFFF}0     & \cellcolor[HTML]{FFFFFF}0     & \cellcolor[HTML]{FFFFFF}2     & \cellcolor[HTML]{FFFFFF}0     & \cellcolor[HTML]{FFFFFF}1     \\ \hline \textbf{Total}          & \cellcolor[HTML]{FFFFFF}\textbf{18,090}      & \cellcolor[HTML]{FFFFFF}\textbf{4,850}     & \cellcolor[HTML]{FFFFFF}\textbf{10,368}     & \cellcolor[HTML]{FFFFFF}\textbf{9,715}     & \cellcolor[HTML]{FFFFFF}\textbf{8,774}     & \cellcolor[HTML]{FFFFFF}\textbf{7,418}     \\ \hline
\end{tabular}
\end{table*}